\documentclass[a4paper,11pt]{article}
\pdfoutput=1 % if your are submitting a pdflatex (i.e. if you have
\usepackage{jcappub} % for details on the use of the package, please
\usepackage[T1]{fontenc} % if needed
\usepackage{amssymb,amsmath}
\usepackage{tikz}
\usepackage{hyperref}

\title{Cosmological Structure Formation in Decaying Dark Matter Models}

\author[1]{Dalong Cheng,}
\author[1]{M.-C. Chu,}
\author[1]{Jiayu Tang}

\affiliation[1]{Department of Physics and Institute of Theoretical Physics, The Chinese University of Hong Kong, \\
 Shatin, Hong Kong SAR, China}

\emailAdd{dlcheng@phy.cuhk.edu.hk}
\emailAdd{mcchu@phy.cuhk.edu.hk}
\emailAdd{jytang@phy.cuhk.edu.hk}
\abstract{
The standard cold dark matter (CDM) model predicts too many and too dense small structures. We consider an alternative model that the dark matter undergoes two-body decays with cosmological lifetime $\tau$ into only one type of massive daughters with non-relativistic recoil velocity $V_k$.  This decaying dark matter model (DDM) can suppress the structure formation below its free-streaming scale at time scale comparable to $\tau$. Comparing with warm dark matter (WDM), DDM can better reduce the small structures while being consistent with high redshfit observations. We study the cosmological structure formation in DDM by performing self-consistent N-body simulations and point out that cosmological simulations are necessary to understand the DDM structures especially on non-linear scales. We propose empirical fitting functions for the DDM suppression of the mass function and the concentration-mass relation, which depend on the decay parameters lifetime $\tau$, recoil velocity $V_k$ and redshift. The fitting functions lead to accurate reconstruction of the the non-linear power transfer function of DDM to CDM in the framework of halo model. Using these results, we set constraints on the DDM parameter space by demanding that DDM does not induce larger suppression than the Lyman-$\alpha$ constrained WDM models. We further generalize and constrain the DDM models to initial conditions with non-trivial mother fractions and show that the halo model predictions are still valid after considering a global decayed fraction. Finally, we point out that the DDM is unlikely to resolve the disagreement on cluster numbers between the Planck primary CMB prediction and the Sunyaev-Zeldovich (SZ) effect number count for $\tau \sim H_{0}^{-1}$.
}

\begin{document}
\maketitle
\flushbottom

\section{Introduction}\label{inro}
Although there are various evidences supporting the existence of dark matter, e.g. the galaxy rotation curves \citep{rotation-curve-1, rotation-curve-2, rotation-curve-3} and bullet clusters \citep{ bullet-cluster-1}, the nature of dark matter is still unknown. Today's concordance cosmology model, which has shown great success in matching with the precision measurements of the cosmic microwave background anisotropies (CMBA) \citep{wmap-1, planck-1} and surveys of the large scale structures (LSS) \citep{bao-1, bao-2},  assumes existence of cold dark matter (CDM). However, observations on galactic and sub-galactic scales have shown conflicts with the CDM predictions. Firstly, high resolution simulations suggest far more dwarf satellites in the local environment than have been observed \citep{missing-1}. Secondly, the CDM haloes have cuspy profiles, which contradicts with the observed shallower profiles of dwarf galaxies \citep{dwarf-1, dwarf-2}. Thirdly, the largest CDM Milky Way satellites have much larger circular velocities to match with observed satellites \citep{too-big-to-fail}, and fourthly the observed galaxy velocity function is significantly lower than the CDM prediction \citep{vf-1, vf-2}.  These problems indicate that the CDM model may generate too many and too dense small structures. Alternative dark matter models, with the property of preserving the large scale virtues of CDM while suppressing small scale structures, are worth considering. In this study, we assume that dark matter undergoes decays. Early discussions of unstable dark matter can be found in Ref. \citep{doro-1, doro-2}.  The decaying dark matter (DDM) can suppress structure formation in general, but the detailed process still depends on the explicit decay models. 

One well studied scenario is the relativistic decay. This model naturally requires the lifetime ($\tau$) to be comparable to cosmic age ($H_0^{-1}$) to preserve dark matter in today's universe. Consequently, haloes would experience adiabatic expansion due to the slow mass loss. With half of the dark matter disappeared, Ref. \citep{ddm-cen} found that the expansion can significantly lower the halo density concentration and prohibit star formation in dwarf satellites. The relativistic products might also interact with the inter-galactic medium and further alter the formation of first structures and the reionization history of the universe \citep{ddm-ripa, ddm-chen, ddm-bie}. However, relativistic decays should change the expansion history of the universe and hence the growth rate of perturbations. The lifetime of DDM in this scenario is tightly constrained by the Integrated Sachs-Wolfe (ISW) effect of the CMBA, so that less than 10\% of the decays could have occurred till now \citep{ddm-ich, ddm-ami}, and the upper limit might become even restrictive from upcoming weak lensing surveys \citep{ddm-wang-1}. 

These constraints do not apply if the decay products keep most mass of their mothers and so remain non-relativistic. The lifetime of this kind can be very short with the decays completed within the radiation domination epoch. The produced daughters can act effectively as warm dark matter (WDM) \citep{ddm-str}, where the linear theory is adequate to understand the effects on perturbations. With sufficient free-streaming length and low phase-space density, the daughters may alleviate the CDM problems \citep{ddm-str, ddm-kap, ddm-cem}, although it is still challenging if they are decayed from thermally produced Weakly Interacting Mass Particles (WIMPs) \citep{ddm-bor}.  Ref. \citep{chou} also argued that the lifetime comparable to the formation of first galaxies may offer an explanation of the ultra-high energy cosmic rays.

Contrary to the short-lifetime decay, we consider non-relativistic decays with long lifetime ($\tau \gtrsim H_0^{-1}$). We will show that such models are completely different from the WDM model. To be more explicit, we consider models in which the mother particle undergoes two-body decay with just one type of massive daughters, where the only two possibilities are
\begin{equation}\label{eq1}
 ddm \rightarrow dm + l 
\end{equation}
and
\begin{equation}\label{eq2}
ddm \rightarrow dm + dm.
\end{equation}
The mother particle $ddm$ is the decaying dark matter and the daughter particle $dm$ is the stable dark matter. They are assumed to be collisionless in the structure formation. In Eq. (\ref{eq1}) (Model A), decay also produces a relativistic daughter $l$, which might be a photon or lepton of the standard model of particle physics. We further assume $l$ to be dark to avoid the constraints from the gamma ray and neutrino measurement of the galactic center and the diffuse background \citep{ddm-bell-1}. This restriction is not needed in Eq. (\ref{eq2}) (Model B). In both situations,  $dm$ receives recoil velocity in the center-of-mass frame of  $ddm$ after production. With a small mass difference $\Delta m \ll m$ between the mother and the total mass of daughters, the recoil velocities ($V_k$) are
\begin{equation}\label{eq3}
 V_k =  c\frac{\Delta m}{m}
\end{equation}
and
\begin{equation}\label{eq4}
 V_k = c\sqrt{2\frac{\Delta m}{m}},
\end{equation}
respectively, where $c$ is the speed of light and $m$ is the mass of $ddm$. We define the lifetime and recoil velocity as the decay parameters and restrict $V_k$ to be less than 1000 km/s for this study. 

The DDM effects within this parameter space automatically mix with the non-linear gravitational evolution \footnote{ See Section \ref{sec3-1}.}, making simple extension of the linear calculation from short to long lifetime inadequate. The induced coupled equations of structure formation have to be solved numerically from the first principle. However, previous numerical studies of this problem have mainly focused on the properties of isolated haloes \citep{ddm-peter-1, ddm-peter-2}. Without the cosmological environment, the growth histories of structures are oversimplified and the statistical information such as the mass function of haloes and matter power spectrum cannot be determined. In this work, we report cosmological simulations of this problem and develop empirical relations to quantify the DDM effects on non-linear scales. 

 We organize the paper as follows:  the equations of the DDM structure formation and the N-body implementation are described in Section \ref{sec2}. In Section \ref{sec3}, we present the features of DDM on structure formation, which are originated separately from the decay parameters.  In Section \ref{sec4}, we model the power spectrum suppression of DDM to CDM in the framework of halo model with unbounded mass. Empirical functions are proposed for the DDM mass function and concentration-mass (c-M) relation.  In Section \ref{sec5}, we discuss the constraints of the DDM parameter space by comparing the power suppression with that of the Lyman-$\alpha$ limited WDM models. In this part, we also relax our implicit assumption that the mother particles completely dominate the dark matter initially. Furthermore, we argue that the DDM suppression is not promising as an explanation of the Planck cluster count disagreement \citep{planck-2013, planck-2015}. Finally, we summarize in Section \ref{sec6}.

\section{The N-body method and simulations}\label{sec2}

\subsection{Equations of the structure formation}\label{sec2-1}
In the non-relativistic situation, $\Delta m / m$ is typically smaller than the order of $10^{-3}$. It is thus a good approximation to set the daughter particle's mass to be the same as the mother's for Model A and half of the mass for Model B. We use the comoving coordinate $\textbf{x}$ and define the particle momentum as
\begin{equation}
 \textbf{p} = a^2m_{x} \frac{\text{d}\textbf{x}}{\text{d} t}, 
\end{equation}
where $a$ is the scale factor and $m_{x}$ is the mass of a particle. The dynamics of the momentum obeys
\begin{equation}
\frac{\text{d} \textbf{p}}{\text{d} t} = -\frac{m_x}{a}\nabla \delta \Phi,
\label{momentum-eq}
\end{equation}
where $\delta \Phi$ is the peculiar potential satisfying 
\begin{equation}
\nabla^2 \delta \Phi = 4\pi G \left[ \rho(\textbf{x}, t) - \bar{\rho}  \right].
\end{equation}

The Boltzmann equation for the distribution function of the mother particle $f_M$ in model A is
\begin{equation}
\frac{\text{d}f_{M}(\textbf{x}, \textbf{p}, t)}{\text{d}t} = -\lambda f_{M}(\textbf{x}, \textbf{p}, t),
\label{boltz-mother}
\end{equation}
where $\lambda = \ln 2 / \tau$ and $\frac{\text{d}}{\text{d} t}= \frac{\text{d}\textbf{x}}{\text{d}t}\cdot \frac{\partial}{\partial \textbf{x}} + \frac{\text{d}\textbf{p}}{\text{d} t}\cdot \frac{\partial}{\partial \textbf{p}}+\frac{\partial}{\partial t}$. The distribution function of the daughter particles satisfies
\begin{equation}
\frac{\text{d}f_{D}^{(1)}(\textbf{x}, \textbf{p}, t)}{\text{d}t} = \int \lambda f_{M}(\textbf{x}, \textbf{p}^{'}, t) \delta_{D}(|\textbf{p}^{'}-\textbf{p}|-amV_k) \frac{1}{A}\text{d}^3\textbf{p}^{'}.
\label{boltz-daughter-1}
\end{equation}
The delta function states the selection rule that the mother should have a velocity difference $V_k$ with its daughter. Because the decays are isotropic in the rest frame of the mother, the constant A is calculable from the number conservation:
\begin{equation}
\int \left[ \frac{\text{d}f_{M}(\textbf{x}, \textbf{p}, t)}{\text{d}t} + \frac{\text{d}f_{D}^{(1)}(\textbf{x}, \textbf{p}, t)}{\text{d}t} \right] \text{d}^3{\textbf{p}} = 0,
\label{boltz-normal-eq-1}
\end{equation}
and we obtain
\begin{equation}
A = 4\pi (amV_k)^2.
\label{boltz-normal-1}
\end{equation}

For Model B, the Boltzmann equation of the mother particles is unchanged. But there are a few modifications for that of the daughter's. Firstly, the selection rule is altered to
\begin{equation}
|\textbf{p}^{'}- 2\textbf{p}| -amV_k = 0,
\label{selection-rule-2}
\end{equation}
as the daughter takes only half of the mother's mass. Besides, the number of produced massive daughters is also doubled in each decay. The equation is then
\begin{equation}
\frac{\text{d}f_{D}^{(2)}(\textbf{x}, \textbf{p}, t)}{\text{d}t} = \int 2\lambda f_{M}(\textbf{x}, \textbf{p}^{'}, t) \delta_{D}(|\textbf{p}^{'}-2\textbf{p}|-amV_k) \frac{1}{B}\text{d}^3\textbf{p}^{'}.
\label{boltz-daughter-2}
\end{equation}
Similarly, the constant $B$ is determined from a modified number conservation relation
\begin{equation}
\int \left[ \frac{\text{d}f_{M}(\textbf{x}, \textbf{p}, t)}{\text{d}t} + \frac{1}{2}\frac{\text{d}f_{D}^{(2)}(\textbf{x}, \textbf{p}, t)}{\text{d}t} \right] \text{d}^3{\textbf{p}} = 0,
\label{boltz-normal-eq-2}
\end{equation} 
leading to
\begin{equation}
B = \frac{\pi (amV_k)^2}{2}.
\label{boltz-normal-2}
\end{equation}

The solutions of the daughter's Boltzmann equations in the two models are actually related.  It is easy to verify that the solution of Eq. (\ref{boltz-daughter-2}) can be expressed using that of Eq. (\ref{boltz-daughter-1}), if
\begin{equation}
f_{D}^{(2)}(\textbf{x}, \textbf{p}, t) = 16f_{D}^{(1)}(\textbf{x}, 2\textbf{p}, t).
\label{daughter-solution-connection}
\end{equation}
The density field of Model B is then
\begin{equation}
\begin{split}
\rho^{(2)}(\textbf{x}, t) &= \int \left[mf_{M}(\textbf{x}, \textbf{p}, t)+\frac{m}{2}f_{D}^{(2)}(\textbf{x}, \textbf{p}, t) \right] \text{d}^3\textbf{p} \\
                                            & = \rho^{(1)}(\textbf{x}, t),
\end{split}
\label{identical-density}
\end{equation}
from which we have shown that Model A and B are actually identical in structure formation given the same decay parameters, although their particle physics are different. We therefore will not distinguish them hereafter.

\subsection{N-body algorithm of DDM simulations}\label{sec2-2}
\begin{figure}[tbp]
\centering
\includegraphics[width=1\textwidth]{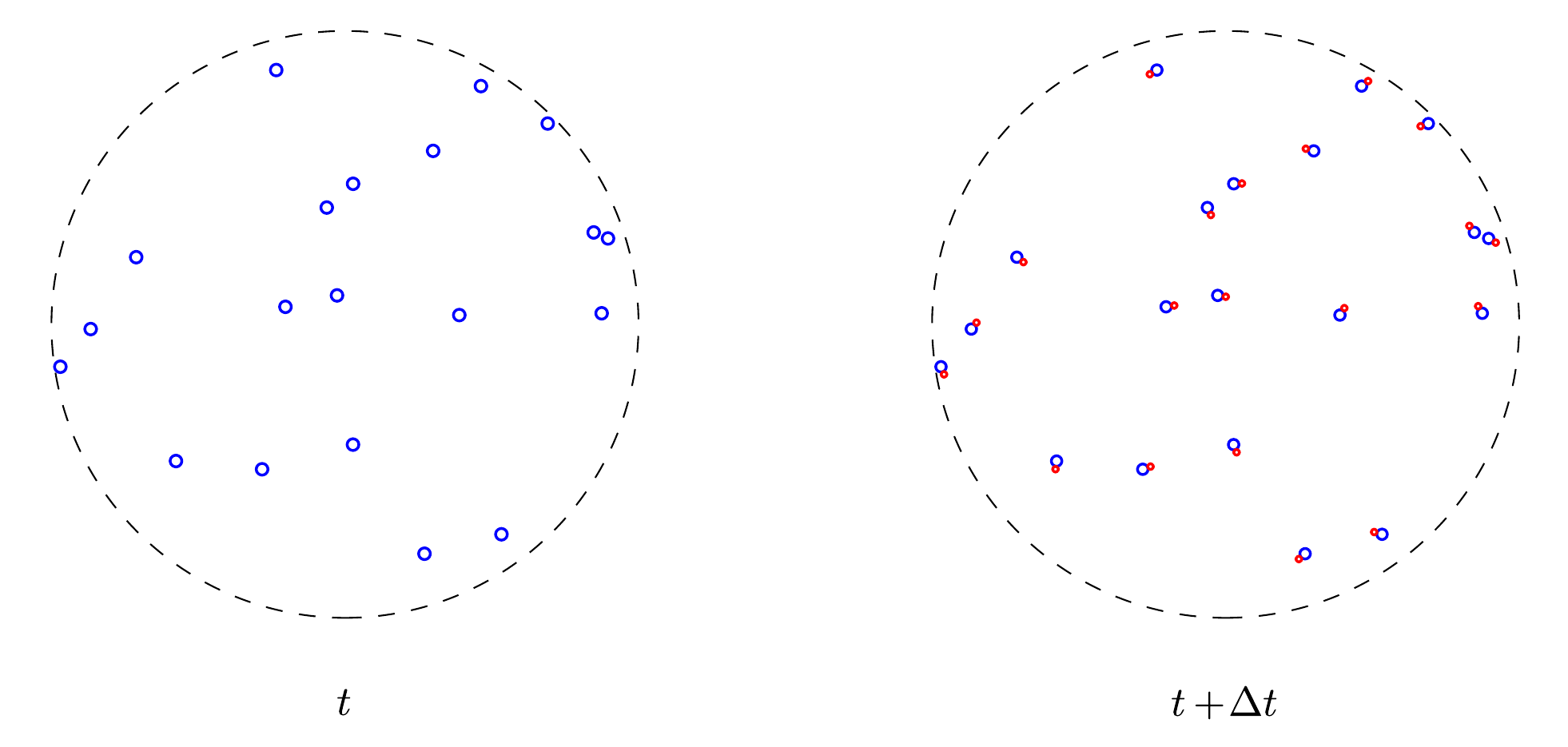}
\caption{An illustration of the N-body description of the DDM models. In the left, we show a sampling of a local density with 20 mother particles at time $t$. They can have very similar momenta if the phase-space element is defined with a narrow $\Delta^3 \textbf{p}$. In the right, we show the daughter simulation particles after a short time of decaying. The daughters are split from the mother particles of the decayed mass and recoiled randomly. The relative positions of the daughters to the mothers represent the directions of the recoil, and we have shown only one daughter simulation particle for each mother. In both plots, the particles are shown in the frame of the mothers. }
\label{fig-ddm-n-body-description}
\end{figure}
N-body simulation is essentially a method to track the underlying phase-space evolution of the matter field. In a local region, a collection of microscopic particles with similar velocities defines a phase-space element \footnote{It equals to $m_{x}f_{x}(\textbf{x}, \textbf{p}, t)\Delta^3\textbf{x}\Delta^3\textbf{p}$, for particles of mass $m_{x}$ in a space interval $\Delta^3\textbf{x}$ and a momentum interval $\Delta^3\textbf{p}$ of the Boltzmann function $f_{x}(\textbf{x}, \textbf{p}, t)$.}, which is then represented by several simulation particles. We illustrate this in the left panel of Fig. \ref{fig-ddm-n-body-description}, where a phase-space element is sampled by 20 simulation particles of similar momenta. In the CDM scenario, the evolution that let each simulation particle follow the dynamic equation Eq. (\ref{momentum-eq}) is equivalent to solving the collisionless Boltzmann equation. In the DDM scenario, microscopic daughter particles after decays leave the original phase-space element of the mothers with a constant recoil velocity to all possible directions. Therefore, sampling the new phase-space elements of the daughters with simulation particles is equivalent to solving the Boltzmann equations of Eq. (\ref{boltz-mother}) and Eq. (\ref{boltz-daughter-1}). In the right panel of Fig. \ref{fig-ddm-n-body-description}, we approach this by letting each simulation particle split the decayed mass to a new simulation particle with a randomly directed velocity. The splitting and new-born particles are noted as the mother and daughter simulation particles. The density field inferred from the daughter simulation is proportional to the density field inferred from the mother particles, representing that the decays of the microscopic particles are proportional to the local density. Also the dispersion of the microscopic daughter particles is represented after considering the splittings of all the mother simulation particles. Notice that the daughter simulation particles represent the mass emitted from the decays, they are not further split in the following evolution.  This N-body description also explains the equivalence of Model A and B, because they generate the same isotropic daughter phase-space elements in the frame of the mothers.

Previously, Peter et al. \citep{ddm-peter-2} proposed an N-body algorithm to study the DDM effects on isolated haloes, which has recently been applied to cosmological structures \citep{ddm-wang-2, ddm-wang-3}. Peter's method preserves the total number of simulation particles and samples new phase-space elements of daughters by randomly choosing and kicking the mother simulation particles according to the decay probability of a time interval. Their algorithm and ours shall be identical for the well sampled phase-space elements of mothers. However, for the badly resolved ones, such as the small structures or the inner regions of haloes, the variance of random picking from the expected value will increase quickly as the number of mother simulation particles drops ($\sigma / \bar{n} \propto 1/\sqrt{N}$). As a result, the lifetime is effectively not uniform in Peter's method, while ours ensures that.

To describe our algorithm, we introduce two additional parameters: the number of daughter simulation particles $N_s$ produced at each split and the number of splittings $f_s$. $N_s$ means how well the velocity dispersion is sampled at a local position and $f_s$ determines the accuracy of updating the decayed mass. For a system evolves to time $T_s$, the split only happens when the decays accumulate to the mass fraction
\begin{equation}\label{eq5}
\eta = 1-\exp \left(- \frac{\ln2}{\tau} \cdot \frac{T_s}{f_s} \right).
\end{equation}
In actual simulations, we only use $f_s$ such that $\eta$ is just a few percent. These additional parameters are artificial. We check the convergence of our simulation over their choices in Appendix \ref{app-a-1} and conclude that $f_s = 10$ and $N_s = 1$ are adequate for our study.

Our N-body method can be generalized to other decay channels and their mixture with branching ratios. For instance, if the decay involves multi-massive daughters as
\begin{equation}
ddm \rightarrow dm_1 + dm_2,
\label{additional-decay-model-1}
\end{equation}
the mother simulation particle needs to split into two types of the daughter simulation particles, and for three-body decays, such as
\begin{equation}
ddm \rightarrow dm_1 + dm_2+l,
\label{additional-decay-model-3}
\end{equation}
simulation can be also be made after adjusting the amplitudes of the recoil velocities to follow certain distributions.  

We implement the method in the public N-body code \textit{Gadget2} \cite{gadget2}. We assign the daughter simulation particles a larger gravitational smoothing length to prevent the two-body relaxations when the lighter daughters are close to the heavy mothers. Other modifications are made in the tree-code to ensure the accuracy of the force with different softening lengths. Unique IDs are also assigned to the daughter simulation particles so that their evolution can be traced.

\subsection{N-body simulations}\label{sec2-3}

\begin{table}[tbp]
\centering
\begin{tabular}{|c|cccccc|}
\hline
Label & $\frac{\tau}{\text{Gyr}}$ & $\frac{V_k}{\text{km/s}}$ & $\frac{\text{L}}{\text{Mpc}/h}$ & $f_s$ & $N_s$ & $\eta/N_s$ \\
\hline
DS-1 & 13.48 & 100  & 50, 20  & 10 & 1 &  6.70\% \\ 
DS-2 & 26.20 & 100  & 50, 20 & 10 & 1 & 3.50\% \\ 
DS-3 & 13.48 & 200  & 100, 50 & 10 & 1 & 6.70\% \\ 
DS-3a & 13.48 & 200 & 50 & 24 & 1 & 2.85\% \\ 
DS-3b & 13.48 & 200 & 50 & 10 & 2  & 3.35\% \\ 
DS-4 & 26.20 & 200  & 100, 50 & 10 & 1   & 3.50\% \\ 
DS-5 & 13.48 & 500  & 256, 100 & 10 & 1  & 6.70\% \\ 
DS-6 & 26.20 & 500  & 256, 100 & 10 & 1  & 3.50\% \\ 
DS-7 & 13.48 & 1000  & 256 & 10 & 1 & 6.70\%  \\ 
DS-8 & 26.20 & 1000  & 256 & 10 & 1  & 3.50\% \\
WS-0.5  &  -    &  -    & 50  & - & -  & - \\
\hline
\end{tabular}
\caption{ Columns from left to right: run label; lifetime of DDM ($\tau$); recoil velocity ($V_k$); simulation box size (L), split frequency ($f_s$); number of daughters at each split ($N_s$) and daughter to mother mass ratio ($\eta/N_s$).  DS-3a and DS-3b are runs with the larger $f_s$ and $N_s$, respectively. Bigger boxes are used for higher recoil velocities in order to capture their larger suppression scales. For each box size, an additional CDM simulation is run without being listed here.}\label{table1}
\end{table}

We assume a flat universe with the cosmological parameters $\Omega_m=0.3$, $\Omega_b = 0.049$, $\Omega_{\Lambda} = 0.7$, $h=0.7$, $n_s=0.96$ and $\sigma_8 = 0.8$, which are in between the WMAP-9yr \cite{wmap-9} and the Planck 2013 results \cite{planck-1}. In the matter dominated epoch, we neglect the baryons and assume DM makes up all the mass fraction of $\Omega_m$ and so do the simulation particles. A correction of the baryon presence will be considered later. The decay parameters are set as $\tau=\{13.48, 26.20\}$ Gyr, which corresponds to $50\%$ or $30\%$ decayed fraction at $z=0$, and $V_k=\{100, 200, 500, 1000\}$ km/s for the DDM simulations.

To highlight the effects of DDM, we also perform CDM and WDM simulations for comparisons. The initial conditions of CDM simulations are generated at $z_i=100$ with $256^3$ simulation particles using the Zeldovich approximation, where we have adopted the BBKS formalism  \citep{bbks} to calculate the transfer function. We consider the simulation box sizes of 256, 100, 50 and 20 Mpc/$h$ with the same number of particles to explore different mass scales. The WDM simulation is set up from the fitting transfer function in Ref. \cite{wdm-aba} with sterile neutrino of 0.5 keV. Because of the long lifetime nature of the DDM models, we neglect the decay effects before $z_i$ \footnote{From example, with half of the DM decayed at $z=0$,  the decayed fractions are just $8.4 \times 10^{-4}$ at $z_i$ and $2.4\times 10^{-5}$ at the decoupling of CMB photons.}, and start the DDM simulations from the same initial conditions as the CDM ones. The DDM simulations also occupy different box sizes according to the recoil velocities under the principle of using smaller box to resolve the effects of smaller $V_k$. We will see in Fig. \ref{fig-ddm-mf-2} and Fig. \ref{fig-ddm-cm} that the DDM effects are consistent in the halo mass functions and profiles over the simulation boxes and thus the mass resolutions. Notice that with respect to the DDM models, there is one more hidden parameter $f_i$ that specifies the mass fraction of $ddm$ in the initial dark matter component. Here, we assume that $ddm$ contributes to all primordial dark matter. We leave further discussion of this parameter in Section \ref{sec5-2}.  More details about the simulations are listed in Table \ref{table1}. The simulations were run on the 72-core cluster of CUHK. With the default choice of the artificial decay parameters, we found that the DDM runs are on average six to eight times slower than the CDM runs, and the ratio can increase linearly with higher $f_s$ and $N_s$.

 To identify haloes, we adopt the density based halo finder \textit{AHF} \cite{ahf}. The halo boundary is defined where the enclosed mean density is 200 times larger than the background critical density $\rho_{crit}$. We further select trusted haloes with more than 100 simulation particles in the CDM and WDM simulations. For the DDM simulations, this requirement is set for the mother simulation particles. We also calculate the power spectrum of the structures using the nearest grid assignment on $1024^3$ grids, where the highest $k$ is truncated at half of the Nyquist frequency to avoid the aliasing effect due to the discrete Fourier transform \cite{dft-jing}. For DDM models simulated with two boxes, we prefer to use the larger box simulations to calculate the DDM power suppression to CDM, where the truncation of long wave modes and the cosmic variance are less important to the power at quasi non-linear regions, which otherwise can cause significant underestimation of the power at $k \sim 1 $ $h/\text{Mpc}$ in simulation box of 20 Mpc/$h$.  In Appendix \ref{app-a-2}, we also test the convergence of the DDM power suppression of DS-3 and DS-4 in box sizes of 100 and 50 Mpc/$h$, and find that the differences are in a few percent level and tend to be even smaller at high $k$ and high redshift.

\section{Features of the DDM structure formation}\label{sec3}
In this section, we show two unique features of the DDM models, which are from the two decay variables.

\subsection{Characteristic suppression scale}\label{sec3-1}

\begin{figure}[!tbp]
\centering
\includegraphics[width=0.7\textwidth]{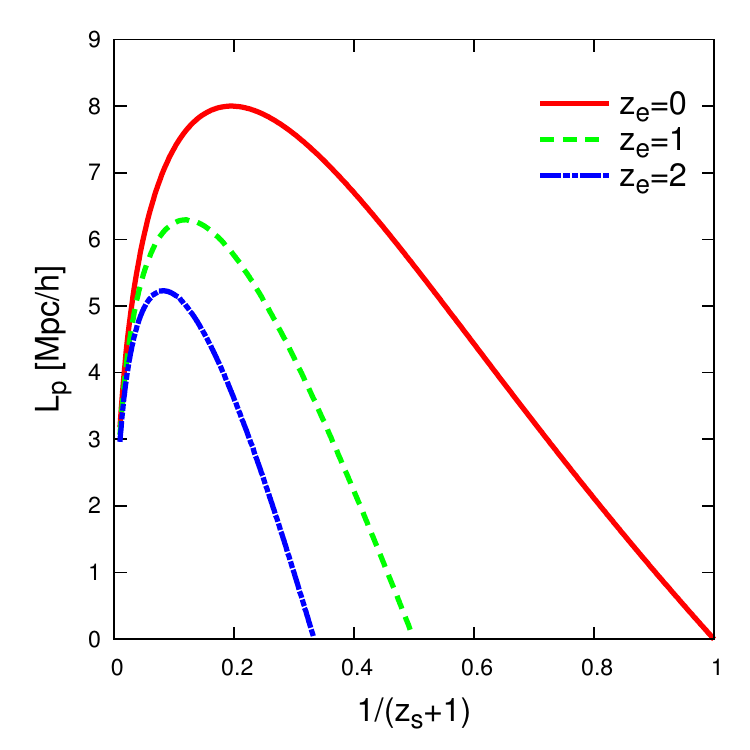}
\caption{ The dependence of comoving propagation distance on the start-redshift when the end-redshift changes. The recoil velocity here is $V_k = 1000$ km/s. The solid (red), dashed (green) and dashed-dotted (blue) lines represent the end-redshift at $z_e =$ 0, 1 and 2, respectively. The recoil velocity here is 1000 km/s. The result can be linearly scaled to other $V_k$. }
\label{fig-free-streaming}
\end{figure}

\begin{figure}[tbp]
\centering
\includegraphics[width=0.7\textwidth]{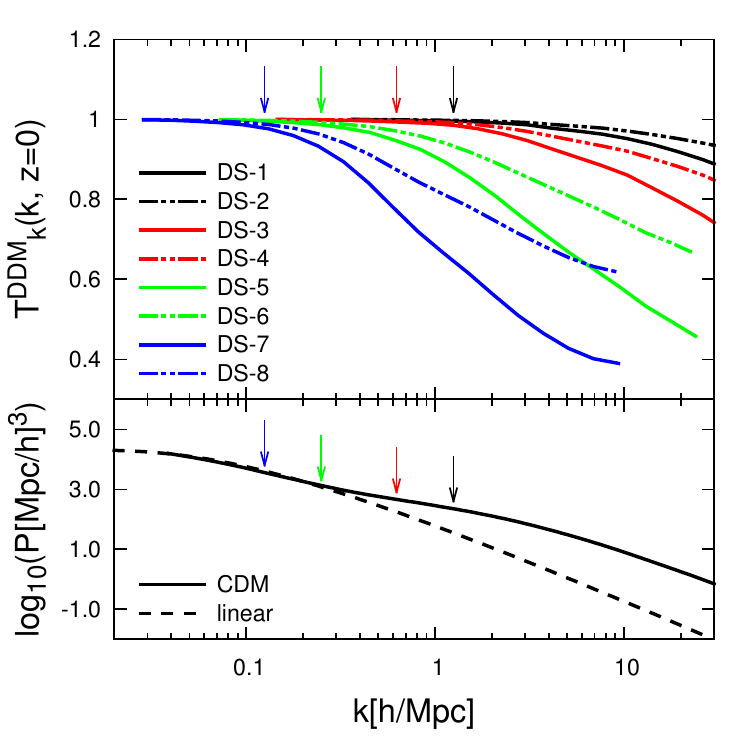}
\caption{\textit{Top:} The transfer functions of DDM to CDM at $z=0$. The colors (black, red, green and blue) denote the recoil velocities (100, 200, 500, and 1000 km/s). The solid and dashed-dotted lines represent different $\tau$ (13.48 and 26.20 Gyr). The arrows point to the scales from Eq. (\ref{eq8}) of each $V_k$.  \textit{Bottom:} The power spectrum of CDM at the same redshift, where the dashed line is the linear power.}
\label{fig-suppression-scales}
\end{figure}

Due to the production of recoiling daughters, the growth of fluctuations shall be suppressed under certain scale. For a daughter produced at $z_s$, the recoil velocity is redshifted as
\begin{equation}\label{eq6}
V(z) = \frac{1+z}{1+z_s}V_k.
\end{equation}
Integrating Eq. (\ref{eq6}) gives us the comoving propagation distance from $z_s$ to $z_e$
\begin{equation}\label{eq7}
L_p = V_k \int_{z_e}^{z_s} \frac{1}{H(z)}\frac{1+z}{1+z_s} dz,
\end{equation}
where $H(z)$ is the Hubble parameter at redshift $z$.  In Fig. \ref{fig-free-streaming}, we plot $L_p$ as functions of $z_s$ for different $z_e$.  With fixed $z_e$, $L_p$ does not increase monotonically with larger $z_s$. Instead, competition exists between the travel time and the redshift of the peculiar velocity. The early produced daughters have more time to travel, but they also experience more redshift in their velocities, while this situation is exactly opposite for the recently produced daughters. As a result,  a maximum value $L_{\text{max}}$ of the propagation can be reached at certain $z_s$, which we define as the free-streaming length for the observer at $z_e$.  We can also see that the free-streaming length always increases with lower $z_e$, because the daughters produced at all redshifts are allowed to travel with more time. 

Similar as Ref. \cite{sup-ks} for WDM, the characteristic scale $k_s$ from which the suppression begins can be estimated as
\begin{equation}
k_s \simeq \frac{1}{L_{\text{max}}},
\label{eq8}
\end{equation}
where $k_s$ is independent of $\tau$. To quantify the relative power spectra of DDM to CDM,  we define the transfer function of DDM as
\begin{equation}
T_k^{ \text{DDM}}(k, z) = \sqrt{\frac{P_{ \text{DDM}}(k, z)}{P_{ \text{CDM}}(k,z)}}.
\label{eq9}
\end{equation}
In the top panel of Fig. \ref{fig-suppression-scales}, we show that the approximation Eq. (\ref{eq8}) agrees well with the simulations, especially for the high recoil velocity cases.  In the bottom panel of Fig. \ref{fig-suppression-scales}, we further mark the suppression scales on the CDM power spectrum at $z=0$. The CDM power begins to enter the non-linear region at $k_{\text{NL}} \simeq 0.3$ $h/\text{Mpc}$. Based on $k_{\text{NL}}$, we divide the DDM simulations into two sets. The small suppression set (DS-1 to DS-4) only modifies the non-linear power, while the large suppression set (DS-5 to DS-8) induces suppression already on linear scales. However, in both sets, the DDM suppression is more significant toward smaller scales, reflecting the importance of including the non-linear evolution for the DDM models.

\subsection{Time evolution}\label{sec3-2}
\begin{figure*}[!htb]
\centering
 \begin{tabular}{@{}ccc@{}}
    \multicolumn{1}{c}{CDM} &
    \multicolumn{1}{c}{DDM} &
    \multicolumn{1}{c}{WDM} \\    
    \includegraphics[trim=100 70 100 70, clip, scale=0.35]{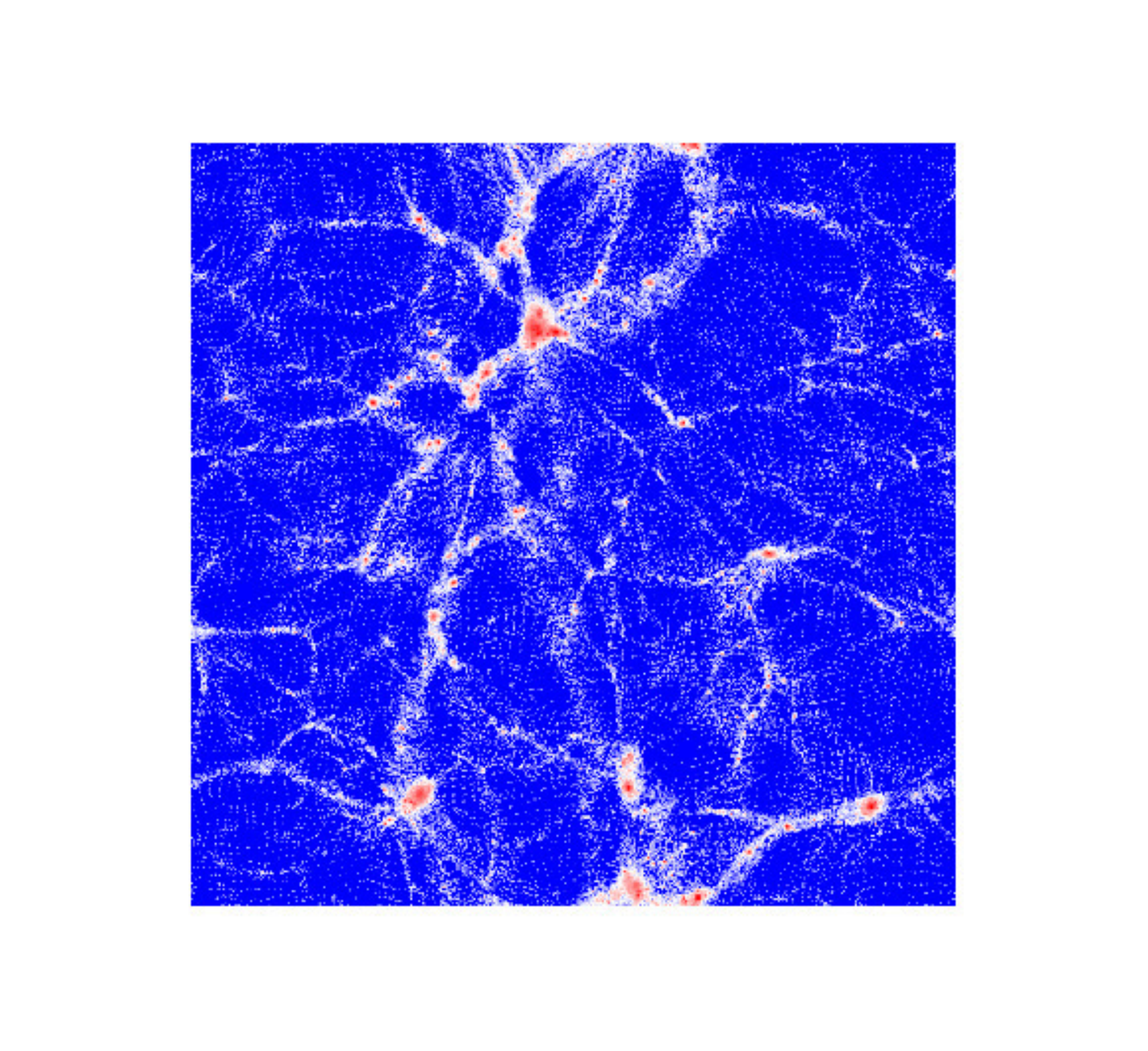}   &
    \includegraphics[trim=100 70 100 70, clip, scale=0.35]{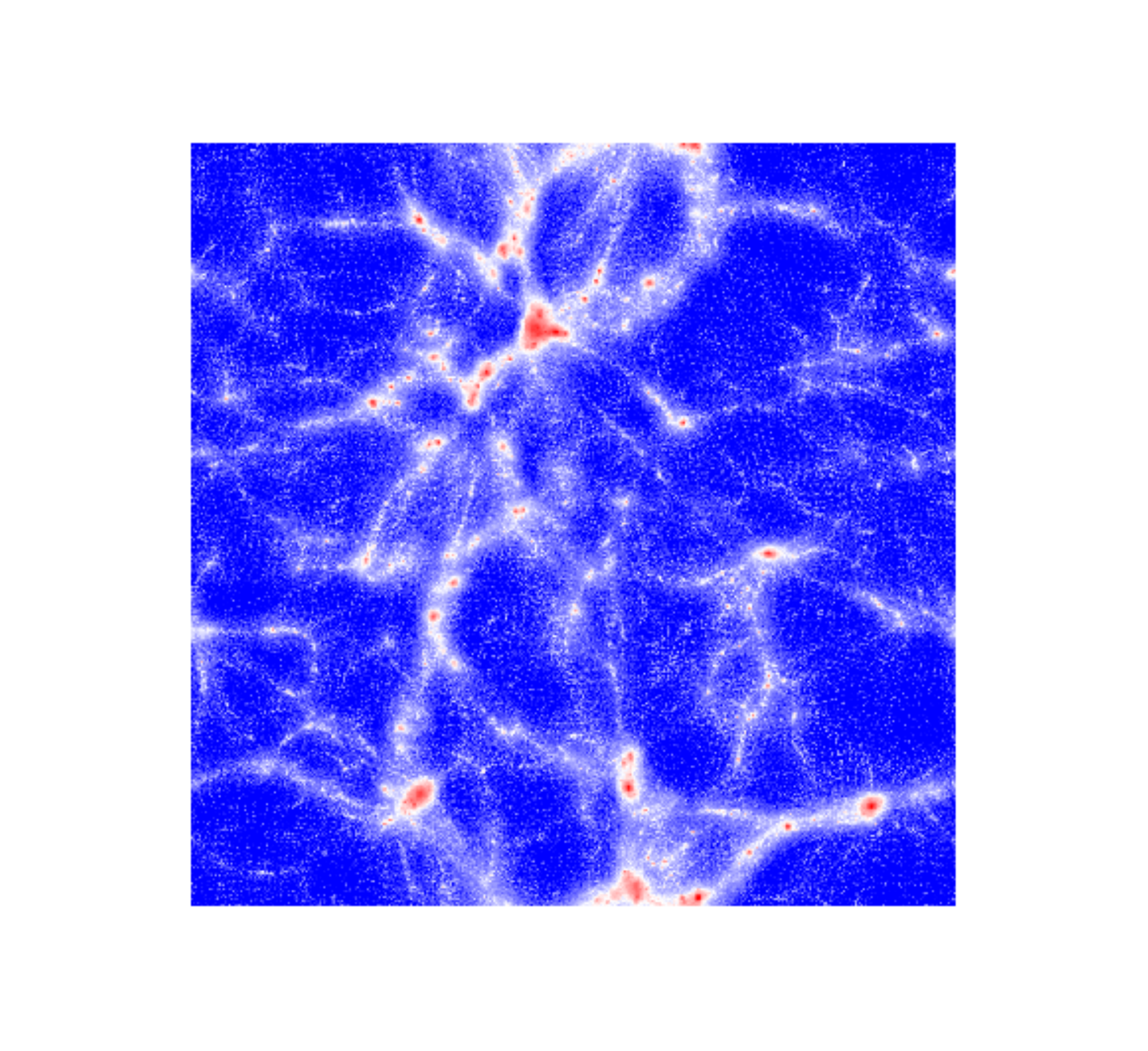}  &
    \includegraphics[trim=100 70 100 70, clip, scale=0.35]{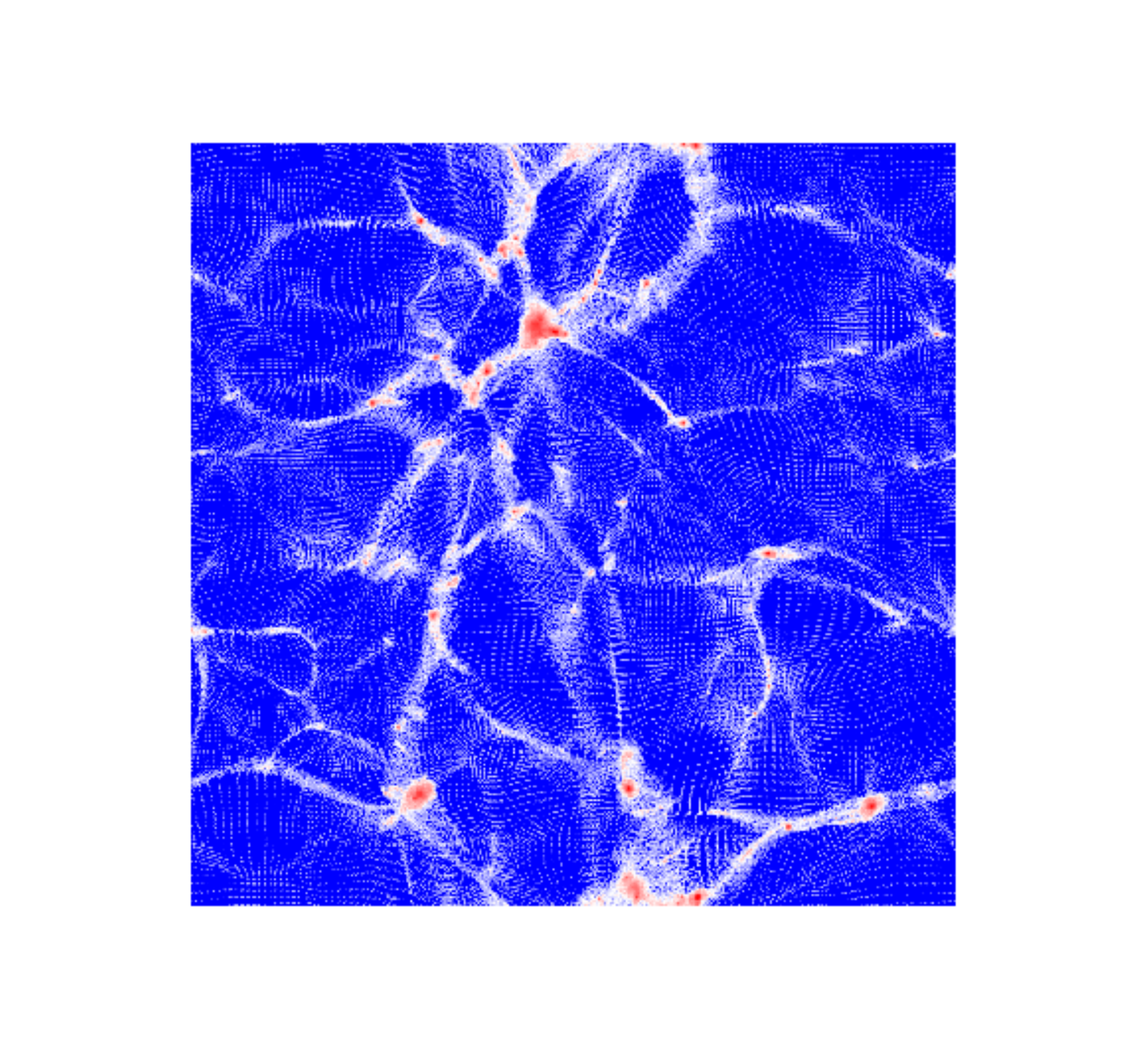}    \\
    \includegraphics[trim=100 70 100 70, clip, scale=0.35]{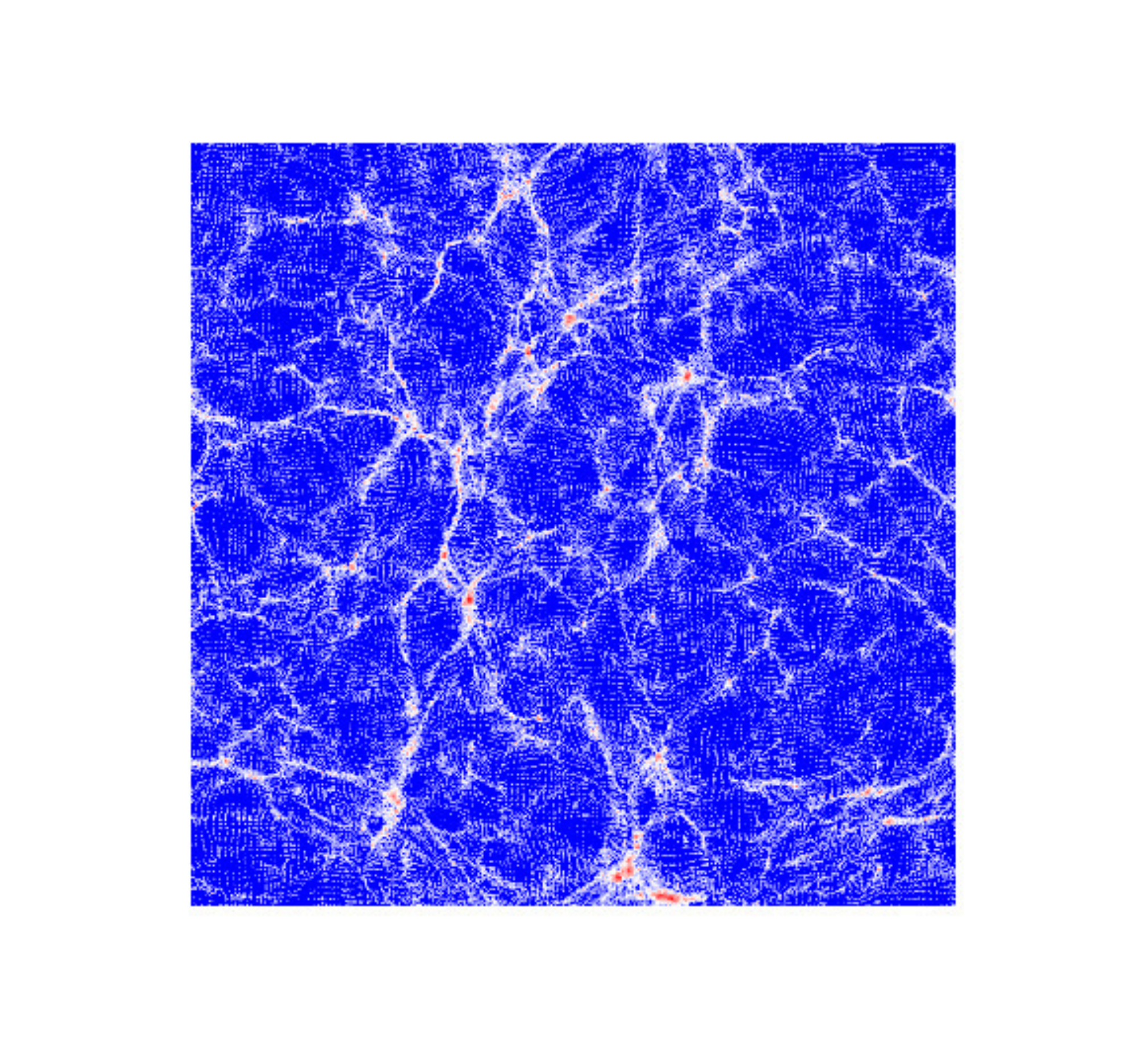} &
    \includegraphics[trim=100 70 100 70, clip, scale=0.35]{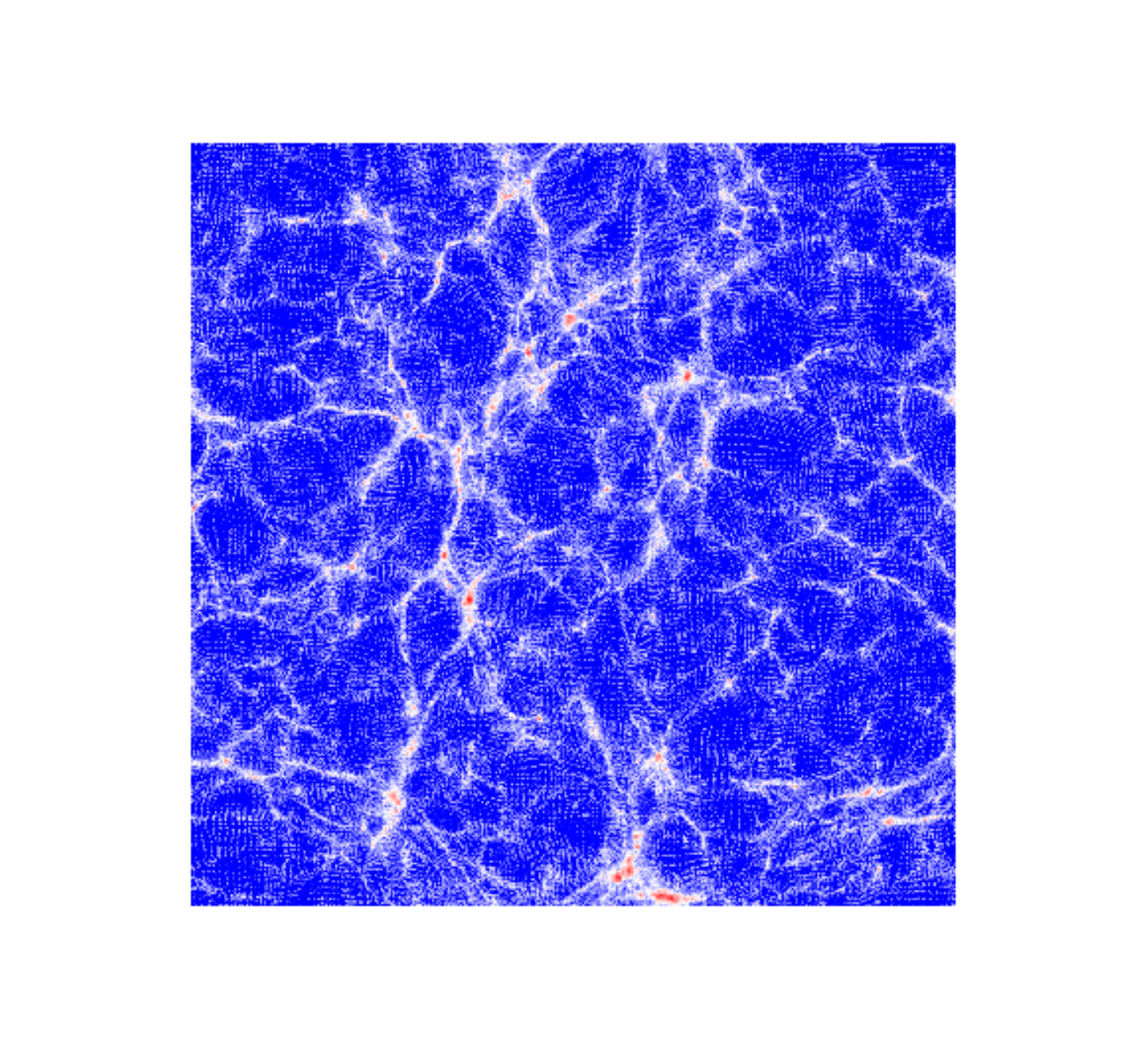}   &    
    \includegraphics[trim=100 70 100 70, clip, scale=0.35]{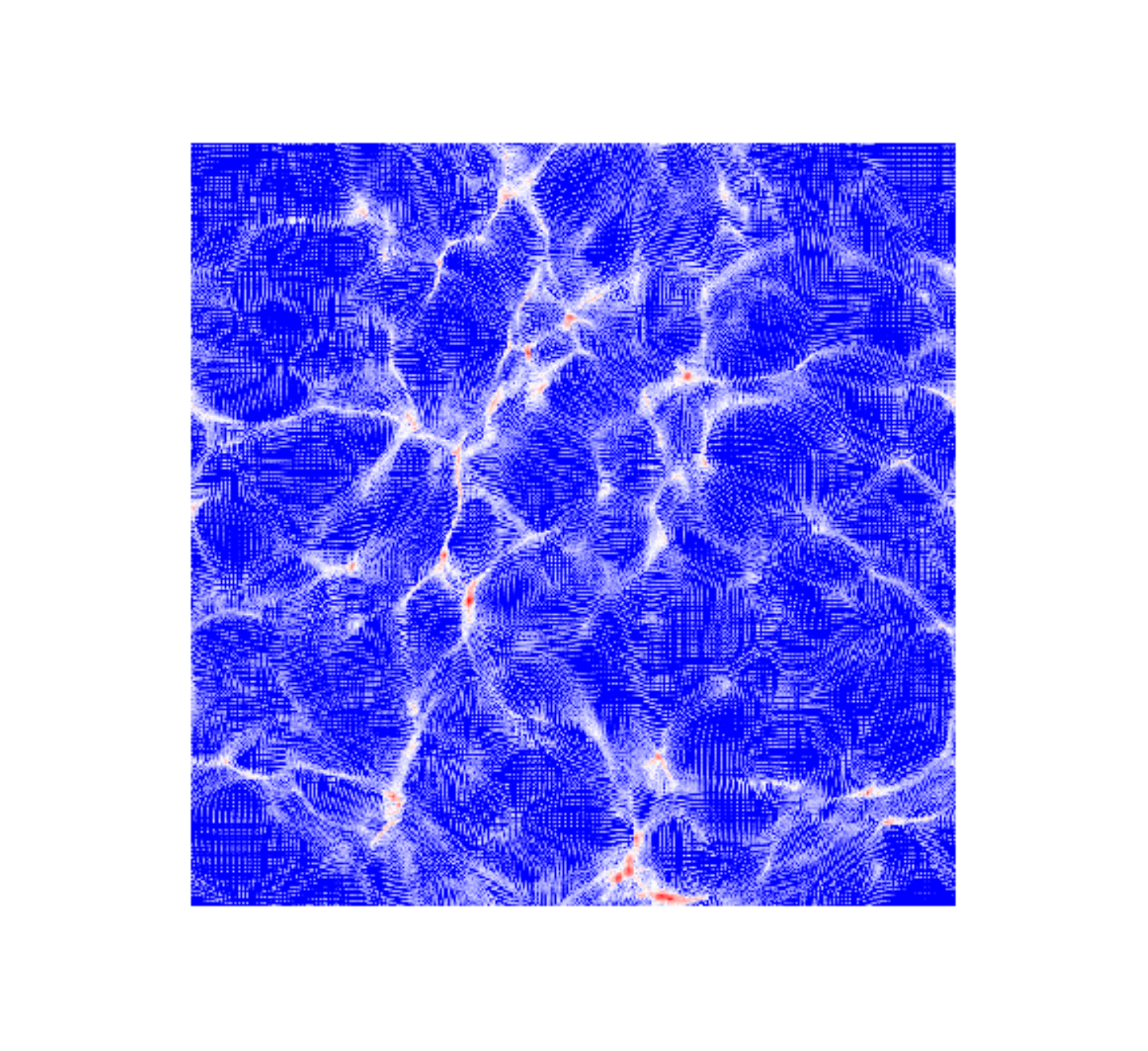}  \\
    \includegraphics[trim=100 70 100 70, clip, scale=0.35]{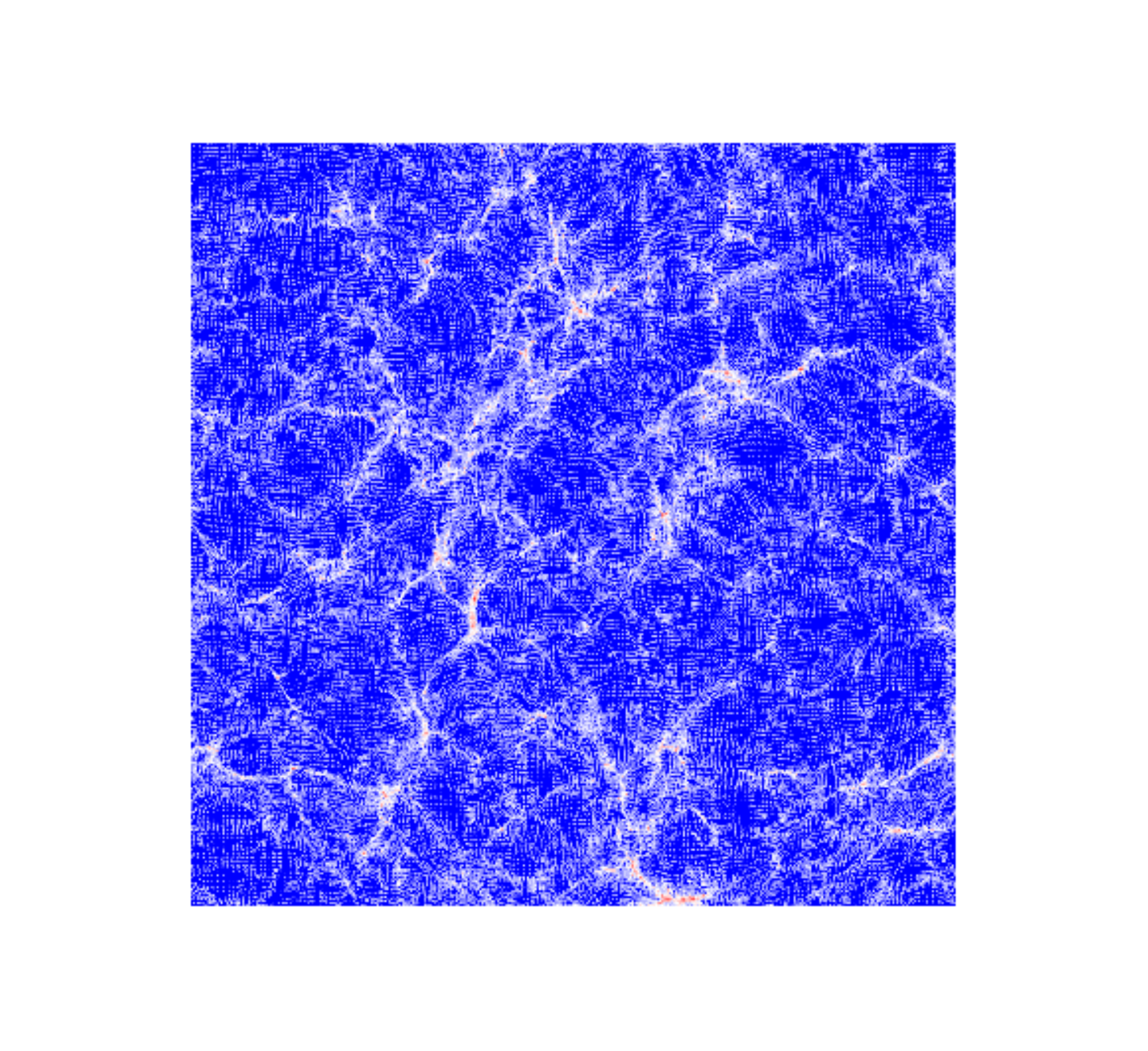} &
    \includegraphics[trim=100 70 100 70, clip, scale=0.35]{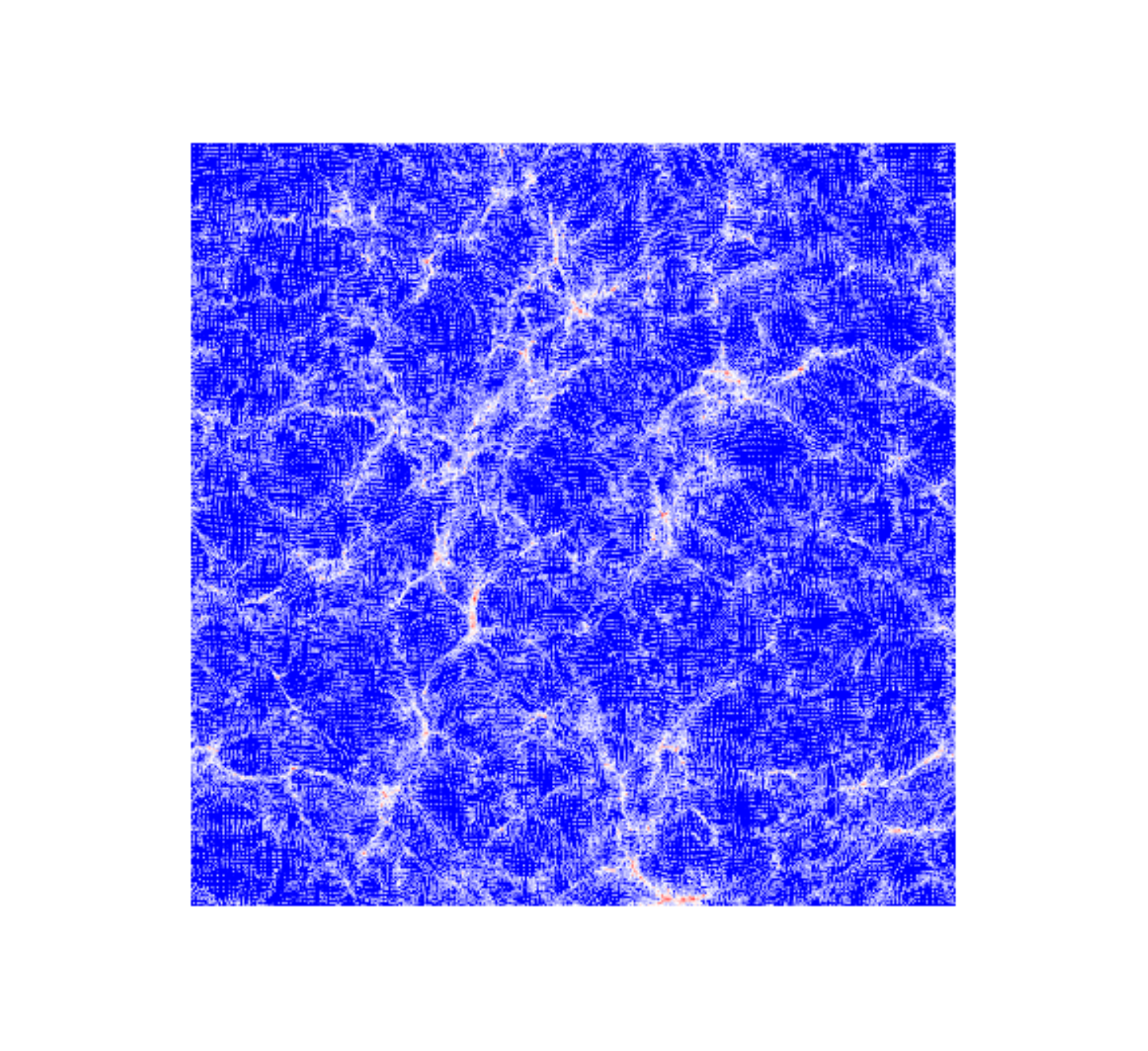}   &    
    \includegraphics[trim=100 70 100 70, clip, scale=0.35]{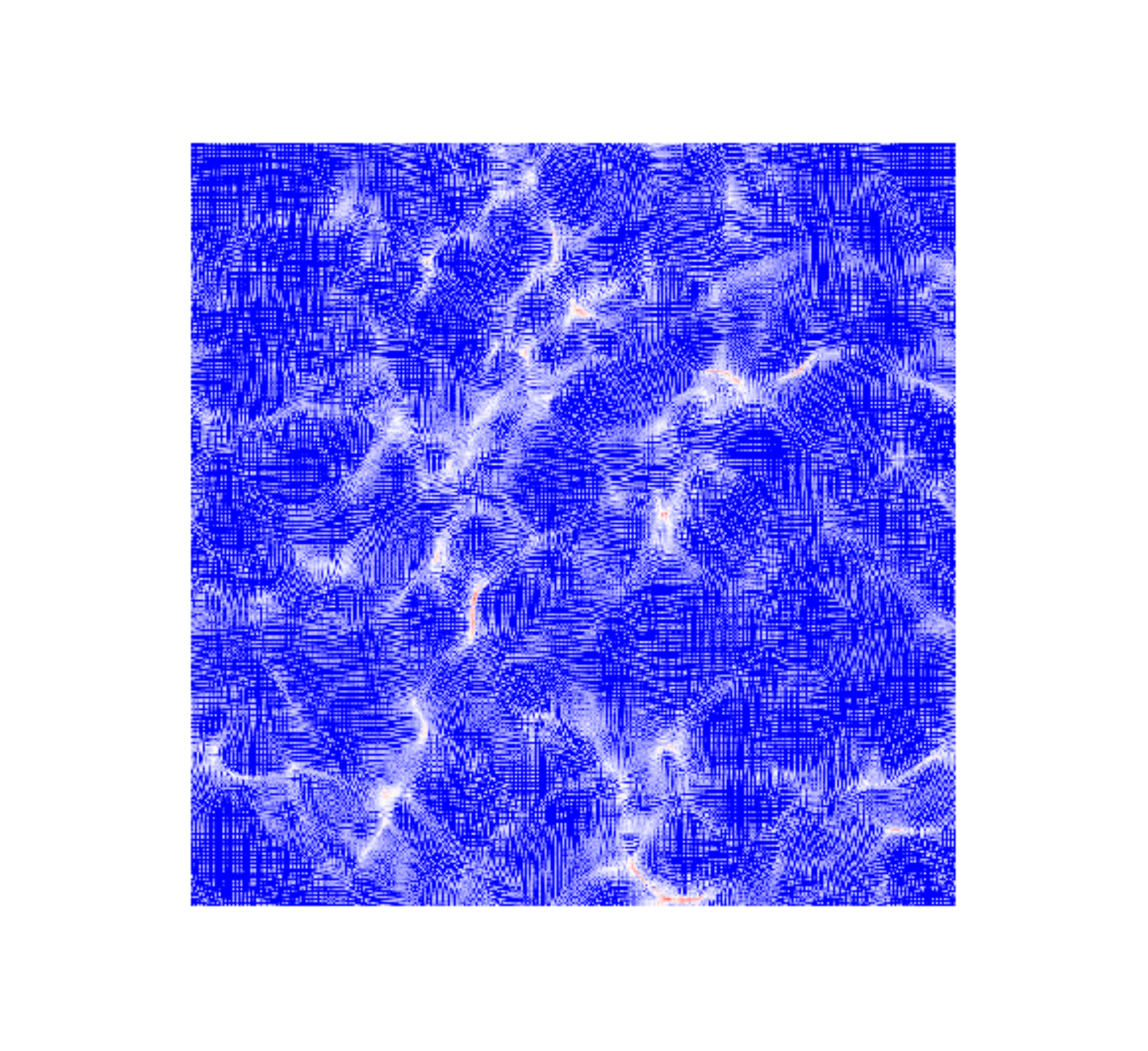}  \\
    \multicolumn{3}{c}{\includegraphics[trim=90 10 100 460, clip, scale=0.5]{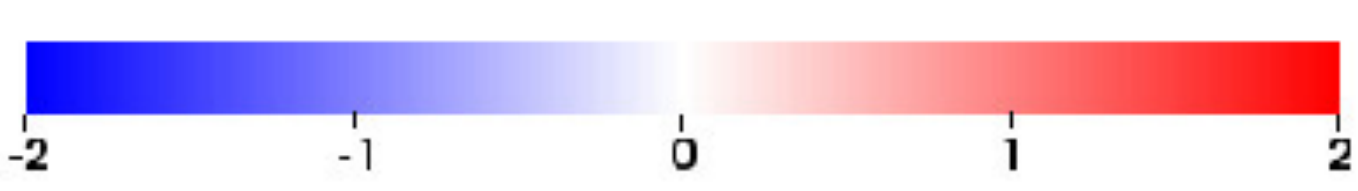}}    \\
    \multicolumn{3}{c}{log$_{10}(1+\delta)$} 
  \end{tabular}
  \caption{Snapshots of 50 Mpc/$h$ width and 10 Mpc/$h$ thickness of the CDM, DDM and WDM simulations. Rows from top to bottom correspond to redshift at $z$=0, 2 and 4. The DDM simulation DS-3 and WDM simulation WD-0.5 are used for the plot.}
\label{fig-cdm-ddm-wdm}
\end{figure*}

\begin{figure}[tbp]
\centering
\includegraphics[width=0.7\textwidth]{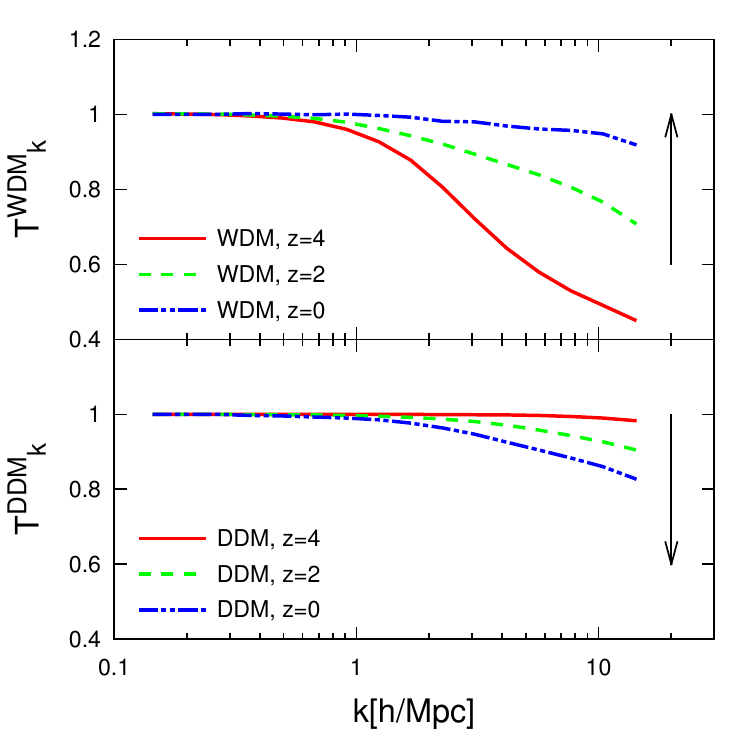}
\caption{ The evolution of the WDM and DDM transfer functions from simulations in Fig. \ref{fig-cdm-ddm-wdm}, shown at $z$=4, 2 and 0 with the solid (red), dashed (green) and dot-dashed (blue) lines respectively. The arrows represent the evolution tendencies from high to low redshift.}
\label{fig-time-evolution}
\end{figure}

Another feature is that the suppression is always larger towards lower redshift as the fraction of decays accumulates. In Fig. \ref{fig-cdm-ddm-wdm}, we show the tendency by comparing the redshift arranged snapshots of the CDM, WDM and DDM simulations. We see that at high redshift the structures are barely discernible between CDM and DDM. At lower redshfit, although the DDM simulation still preserves the overall scheme of the filamentary structures as CDM, the dense regions are quite extended. In contrast, the WDM structures appears to differ from the CDM's mostly at high redshift. To better show the differences, we plot the DDM and WDM power transfer functions relative to CDM in Fig. \ref{fig-time-evolution}. As expected, we observe more suppression in the DDM simulation at lower redshift, which is opposite to the regeneration of small scale powers in WDM \citep{wdm-reg-1, wdm-reg-2, wdm-reg-3}. This opposite evolution is intrinsic to the models and should be useful in distinguishing the DDM and WDM models. Besides, the high reionization redshift ($z \sim 6$) inferred from the quasars absorption lines \cite{agn-reion} requires sufficient small scale fluctuations at high redshift. Being more like the CDM model in the early epoch, the long-lifetime DDM models could more easily be consistent with these observations without compromising the suppression at lower redshift. In contrast, reionization alone has set considerably stringent constraints on the mass of WDM particles \cite{wdm-reion-1, wdm-reion-2, wdm-reion-3}.

\section{Modelling the DDM suppression}\label{sec4}
\subsection{The halo model of DDM}\label{sec4-1}
We try to reconstruct the non-linear DDM transfer functions using halo model. The standard halo model assumes  all matter in the universe in form of haloes. To calculate the power spectrum, the number density,  spatial distribution and profiles of haloes need to be given (see Ref. \citep{hm-1} for a detailed review). However, this assumption does not apply for models that can suppress the formation of small haloes, such as WDM and DDM of this study, due to the fact that there is always unbounded mass. To deal with this problem, the halo model was extended in Ref. \citep{hm-2}. We follow their method and briefly discuss the halo model of this kind. 

The idea is to separate the density field into two parts,
\begin{equation}\label{eq10}
\rho_m(x) = \rho_h(x) + \rho_s(x),
\end{equation}
where $\rho_h$ and $\rho_s$ are the densities of halo and smooth mass. Averaging over the volume, the mean density is $\bar{\rho}_m = \bar{\rho}_h + \bar{\rho}_s$. The halo contribution to the average density is related to its mass function as
\begin{equation}\label{eq11}
\bar{\rho}_h = \int_{M_{\text{ cut}}}^{\infty} dM n(M)M,
\end{equation}
where $n(M) = \text{d} N(>M) / \text{d}  M$ is the halo mass function and $M_{ \text{cut}}$ is a cutoff mass below which haloes are expected not to exist. The halo mass fraction and the density contrasts of the two components are defined as
\begin{equation}\label{eq12}
f=\bar{\rho}_h/\bar{\rho}_m,
\end{equation}
and
\begin{equation}\label{eq13}
\delta_{\chi} = \frac{\rho_{\chi} - \bar{\rho}_{\chi}}{\bar{\rho}_{\chi}},
\end{equation}
where $\chi = \{h, s\}$ stands for the halo or the smooth mass. The total density contrast is then
\begin{equation}\label{eq14}
\delta = f\delta_{h}+(1-f)\delta_{s}.
\end{equation}
In the statistically homogeneous and isotropic universe, the power spectrum can be expressed as
\begin{equation}\label{eq15}
P_{\delta\delta}(k) = (1-f)^2P_{ss}(k) + 2f(1-f)P_{sh}(k) + f^2P_{hh}(k).
\end{equation}
The halo powers are decomposed into the normal one- and two-halo terms
\begin{equation}\label{eq16}
 P_{hh}(k)= P_{1h}(k) + P_{2h}(k),
\end{equation}
with the explicit form
\begin{equation}\label{eq17}
\begin{split}
& P_{1h}(k) = \frac{1}{\bar{\rho}_h^2}\int_{M_{ \text{cut}}}^{\infty} dM n(M) M^2 \tilde{u}^2(k|M), \\
& P_{2h}(k) = \frac{P_{\text{lin}}(k)}{\bar{\rho}_h^2}\left[\int_{M_{ \text{cut}}}^{\infty}dM M b_1(M)n(M)\tilde{u}(k|M)\right]^2,
\end{split}
\end{equation}
where $\tilde{u}(k|M)$ is the Fourier transform of the mass normalized halo profile, $b_1(M)$ is the linear bias for halo of mass $M$, and $P_{\text{lin}}(k)$ is the linear power spectrum. The smooth mass is mapped to the underlying density field with a constant bias as $\delta_s \sim b_s\delta$ and is assumed to correlate with itself and the halo density linearly. The smooth-smooth and smooth-halo terms are then
\begin{equation}\label{eq18}
P_{ss}(k) = b_s^2 P_{\text{lin}}(k),
\end{equation}
and 
\begin{equation}\label{eq19}
P_{sh}(k) = \frac{b_s P_{\text{lin}}(k)}{\bar{\rho}_h} \int_{M_{ \text{cut}}}^{\infty}dM M b_1(M) n(M) \tilde{u}(k|M).
\end{equation}
The bias of the smooth matter is actually not a free parameter but shall be constrained by Eq. (\ref{eq14}), since the density contrast from haloes can also be expressed as
\begin{equation}\label{eq20}
\delta_h = \frac{1}{\bar{\rho}_h}\int_{M_{ \text{cut}}}^{\infty} dM M n(M)b_1(M) \delta,
\end{equation}
at large scales.
Substituting Eq. (\ref{eq20}) back in Eq. (\ref{eq14}), we thus obtain
\begin{equation}\label{eq21}
b_s = \frac{1-f b_{ \text{eff}}}{1-f},
\end{equation}
where we have introduced the effective bias
\begin{equation}\label{eq22}
b_{ \text{eff}} = \frac{1}{\bar{\rho}_h}\int_{M_{ \text{cut}}}^{\infty} dM M n(M)b_1(M).
\end{equation}
To use the model, we need to understand the mass function, halo profiles and linear-bias of the DDM haloes at first. We examine these ingredients in the following one by one.

\subsection{The mass function}\label{sec4-2}
\begin{figure}[tbp]
\centering
\includegraphics[width=0.7\textwidth]{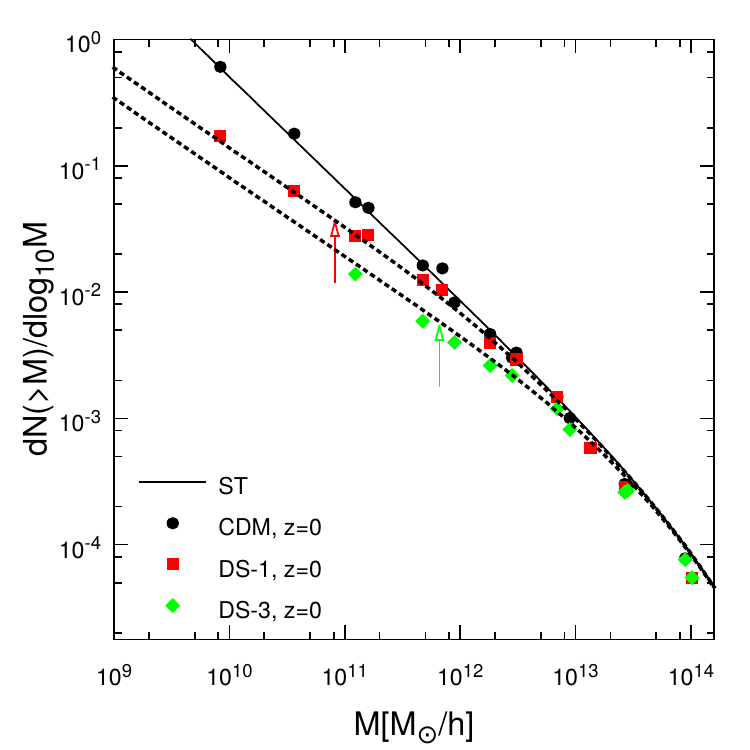}
\caption{The mass functions of the ST formalism (solid line) and measurements from CDM (round black points) and two DDM simulations (square and diamond points).  The CDM data points are measured in simulation boxes of 100, 50 and 20 Mpc/$h$, while data points of DS-1 and DS-3 are obtained by combing the two simulation boxes in Table \ref{table1}.  The dashed lines are following the best fit of Eq. (\ref{eq28}) with the arrows pointing to the cutoff mass indicated in Peter et al. \cite{ddm-peter-2} for the two sets of decay parameters.}
\label{fig-ddm-mf-1}
\end{figure}

\begin{figure}[tbp]
\centering
\includegraphics[width=1\textwidth]{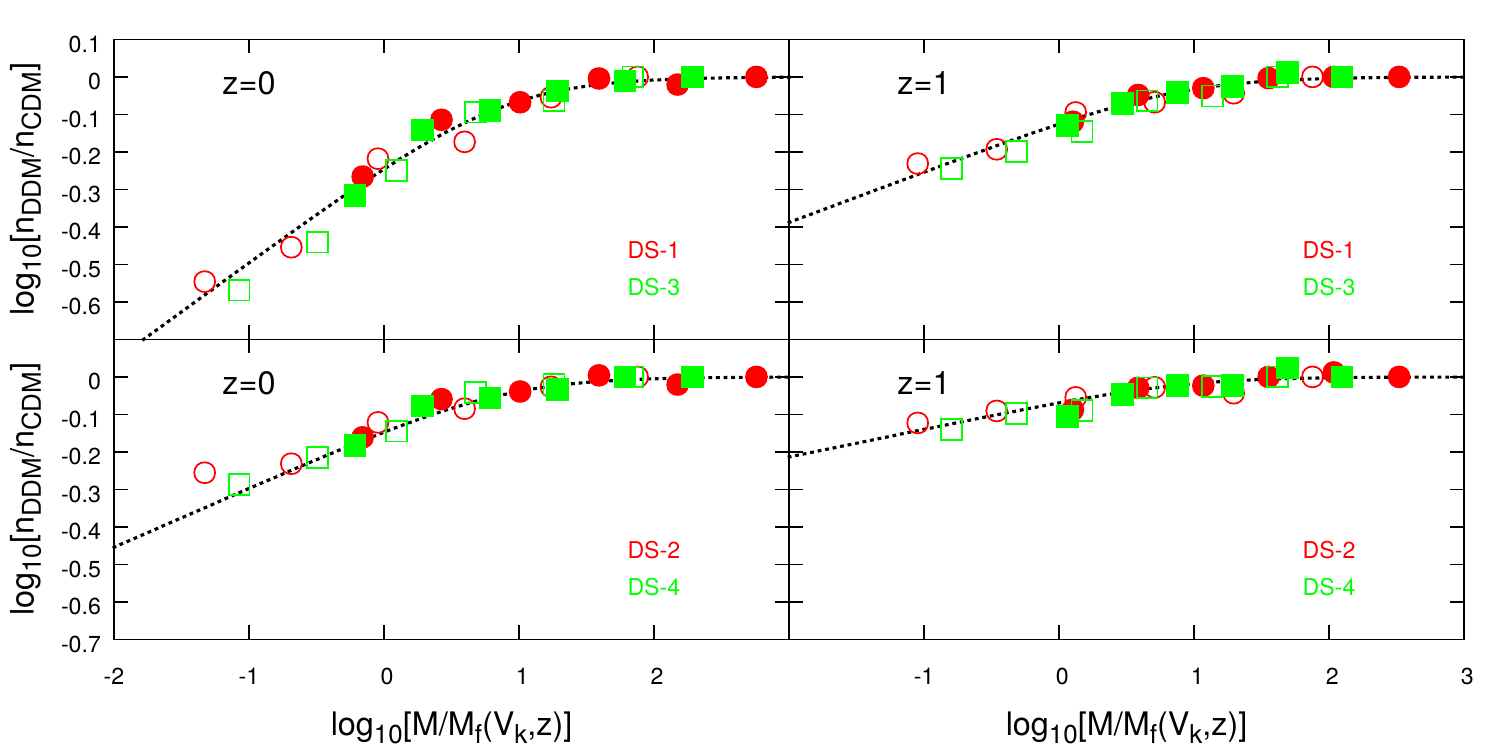}
\caption{The mass function ratio of DDM to CDM as a function of the $M_f$ normalized halo mass for different DDM parameters and redshifts. The dash lines represent the best fit of Eq. (\ref{eq28}). The filled points refer to the measurements in larger boxes, and the not-filled points are from the smaller boxes as in Table \ref{table1} for DS-1 to DS-4, where consistency of simulation boxes and resolutions can be observed.}
\label{fig-ddm-mf-2}
\end{figure}

The mass function of CDM is well studied with the excursion set theory \cite{ps-mf}, where the number density of haloes is related to the appearance probability of density peaks in the halo patch averaged over all ensembles. The usual parameterization of the mass function is
\begin{equation}\label{eq23}
n(M)=\frac{1}{2}\frac{\bar{\rho}_m}{M^2}f(\nu)\left|\frac{\text{d} \log \sigma^2}{\text{d} \log M}\right|;   \nu = \frac{\delta_c(z)}{\sigma(M)},
\end{equation}
where $\bar{\rho}_m$ is the average matter density, $\delta_c(z) = 1.686/D(z)$ is the collapse threshold and $D(z)$ is the linear growth factor. The variance of the overdensity at radius $R = (3M/4\pi\bar{\rho}_m)^{-1/3}$ is defined as
\begin{equation}\label{eq24}
\sigma^2(M) = \int \frac{d^3 \textbf{k}}{(2\pi)^3}P_{\text{lin}}(k)W^2(kR),
\end{equation}
where $W(y) = 3(\sin y - y\cos y)/y^3$ is the Fourier transform of the top-hat windows function.  We adapt the ST formalism \cite{st-mf} for CDM 
\begin{equation}\label{eq25}
f(\nu) = A \sqrt{\frac{2}{\pi}}\sqrt{q}\nu\left[1+(\sqrt{q}\nu)^{-2p}\right] \exp \left(-\frac{q\nu^2}{2}\right),
\end{equation}
with $p=0.3$, $q = 0.707$ and the normalization parameter $A = 0.3222$.

Fig. \ref{fig-ddm-mf-1} compares the ST formalism with the measured mass functions of CDM and DDM simulations. The ST formalism shows good agreement with the data of CDM. But for DDM, there is clear suppression below certain mass. DS-3 here has larger recoil velocity than DS-1, and so it deviates from CDM at higher mass. We can also check the predictions of previous isolated studies, where the haloes having the escape velocities equalling to recoil velocities have the mass of $8.2 \times 10^{10}$ and $6.6 \times 10^{11}$ $\text{M}_{\odot}/h$ for $V_k = 100, 200$ km/s, respectively. Different from the suggestion of Peter et al. \cite{ddm-peter-2} that haloes with escape velocity smaller than $V_k$ should be destroyed, we observe no truncation of the DDM halo mass function below these scales in cosmological simulations. Oversimplification of the formation history in isolated studies shall be the reason of the difference. Since small haloes in cosmological simulation are formed earlier, they can survive the decays through adiabatic mass loss but without being completely destructed. Their mergers would still hierarchically form bigger haloes.

To describe the mass function of our simulations, we develop a fitting function in the form 
\begin{equation}\label{eq28}
\frac{n_{ \text{DDM}}(M, z)}{n_{ \text{CDM}}(M,z)} = \left(1+\beta_{m} \frac{M_{f}}{M}\right)^{-\alpha_{m} f_{d}},
\end{equation}
in which $\alpha_m$ and $\beta_m$ are fitting parameters. Here, we have introduced two effective variables
\begin{equation}\label{eq26}
M_{f}= \frac{4\pi}{3}\bar{\rho}_mL^3_{\text{max}}(V_k, z) 
\end{equation}
and
\begin{equation}\label{eq27}
f_{d} = 1- \exp\left[-\frac{\ln 2}{\tau}T(z)\right], 
\end{equation}
where $M_{f}$ is the characteristic mass of the free-streaming length $L_{\text{max}}$ and $f_{d}$ is the decayed fraction at time $T$ of redshift $z$. They are designed to separate the dependence of the suppression on the decay parameters. Our fitting function does not have explicit dependence on the redshift, whose influence is already embedded in $M_f$ and $f_d$.

Fig. \ref{fig-ddm-mf-2} shows the mass function ratio of DDM to CDM versus the $M_f$ normalized halo mass. The compared DDM simulations in each panel have the same $f_d$ but different $M_f$.  The dashed lines are the best fit of Eq. (\ref{eq28}) with the parameters: $\alpha_m = 0.526$ and $\beta_m = 7.61$. The fitting function shows good agreement with simulations of different combinations of decay parameters as well as redshift. Here, we have only considered the small suppression set. The large suppression set is found to deviate significantly from this result by affecting already the linear power. Currently, there is no theoretical work on the mass function of DDM models. Given the simplicity of this empirical approach, our result shall motivate further theoretical considerations on the mass function of DDM models, for example a both scale- and time-dependent barrier in the excursive set theory.  

\subsection{The halo profile}\label{sec4-3}

\begin{figure}[tbp]
\centering
\includegraphics[width=0.7\textwidth]{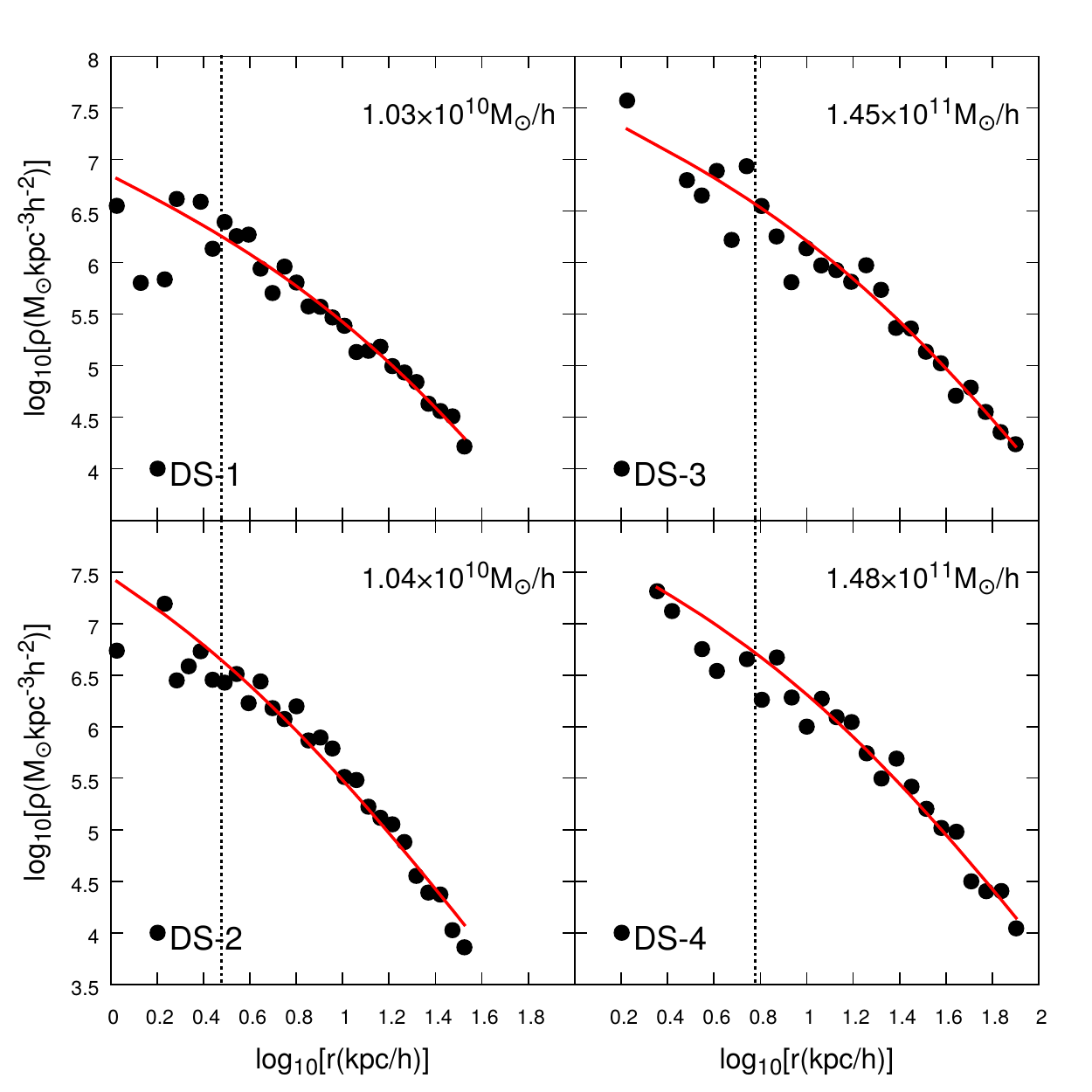}
\caption{The dwarf halo profiles randomly selected from simulations DS-1 to DS-4 at $z=0$. The solid (red) lines are the best-fit of NFW and the vertical dash line marks the spatial resolution of each simulation. }
\label{fig-dwarf-density}
\end{figure}

\begin{figure}[tbp]
\centering
\includegraphics[width=1\textwidth]{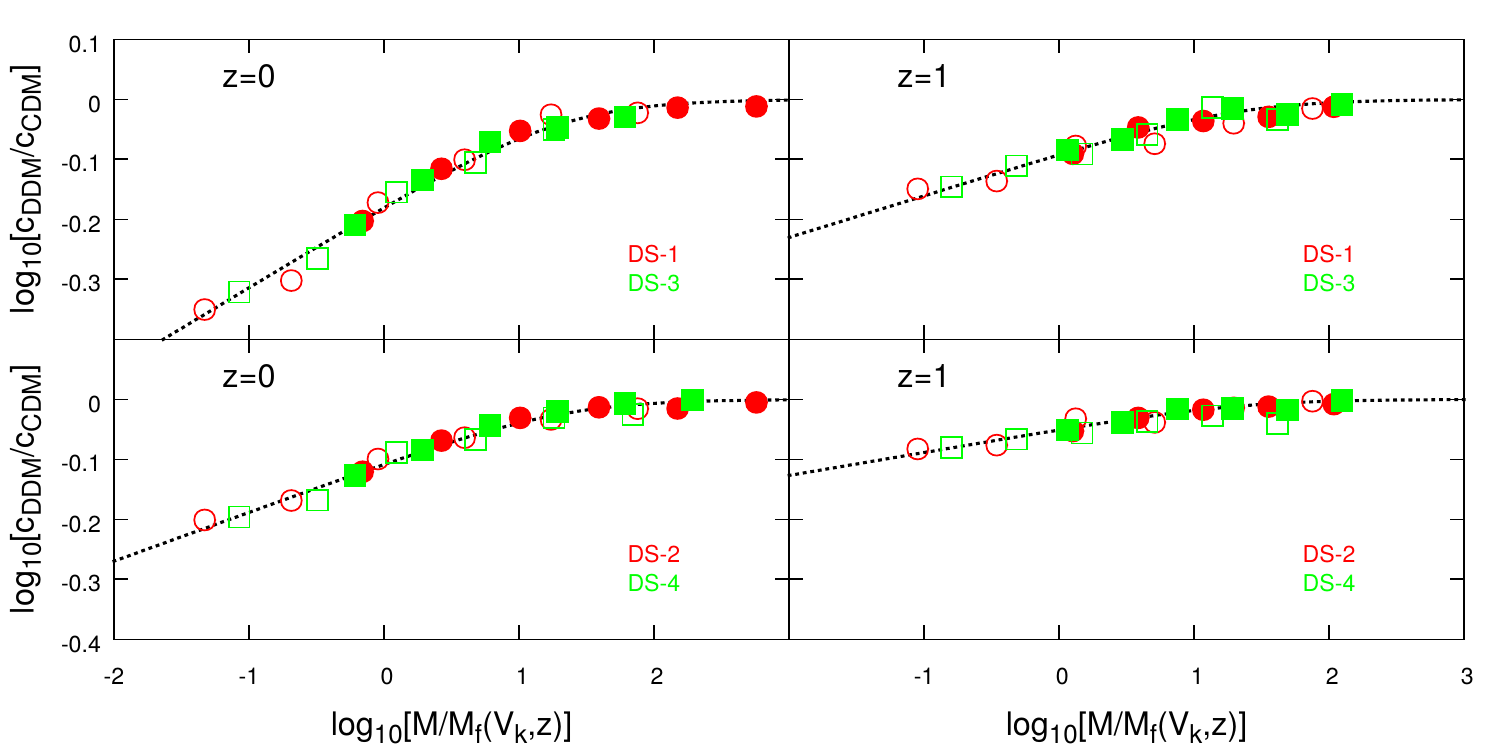}
\caption{Ratio of the NFW concentration between DDM and CDM as a function of normalized mass. The dash lines are the best-fit of Eq. (\ref{eq31}). The point styles are the same as in Fig. \ref{fig-ddm-mf-2}. Consistency of the DDM halo concentration can also be observed over the simulation boxes.}
\label{fig-ddm-cm}
\end{figure}

The density distribution of a CDM halo is usually modelled by the Navarro-Frenk-White (NFW) profile \cite{nfw-1, nfw-2}
\begin{equation}\label{eq29}
\rho(r) = \frac{\rho_{s}r_s}{r(1+r/r_s)^2},
\end{equation}
where $\rho_s$ and $r_s$ are the normalization density and characteristic length scale. It is usually convenient to reparameterize the profile by two other parameters,  the virial mass $M$ and the halo concentration $c=r_{vir}/r_s$, where the virial radius $r_{vir}$ is related to the virial mass through $M = 4\pi/3r_{vir}^3\Delta\rho_{crit}$ with $\Delta =200$ for our halo definition. These two parameters are related after integrating the profile to virial radius, which gives the relation
\begin{equation}\label{eq30}
M = 4\pi\rho_s r_s^3\left[\ln(1+c) -\frac{c}{1+c}\right].
\end{equation}  

In DDM models, haloes are kinetically heated from the recoiling daughters. Previous semi-analytic study expected the formation of cores in dwarf haloes \cite{sanchez-salcedo}. In Fig. \ref{fig-dwarf-density}, we randomly select the dwarf mass range haloes in simulations DS-1 to DS-4 to test this prediction. We show the best-fit NFW profiles with the red lines, and used the vertical dashed lines to present the spatial resolutions of the simulations.  We observe no evidence of core developing within our resolution limits. Obviously, to determine whether there are cores or how they depend on the decay parameters is beyond the scope of these simulations. We leave these questions to future high resolution studies. In this paper, we still stick to the NFW profile for DDM haloes; however, we examine how the concentration changes with the DDM parameters as a quantification of the halo inner density decrease. Similar to the mass function, we parameterize the DDM concentration as
\begin{equation}\label{eq31}
\frac{c_{ \text{DDM}}(M,z)}{c_{ \text{CDM}}(M,z)}=\left(1+\beta_c \frac{M_f}{M}\right)^{-\alpha_c f_d}.
\end{equation}
In Fig. \ref{fig-ddm-cm}, we show the best-fit function with $\alpha_c = 0.271$ and $\beta_c = 20.4$. The decays make little change to the massive haloes ($M \gg M_f$) because of their deep gravitational potentials. While in the low-mass limit, the DDM concentration-mass relation (c-M relation) scales as $c \propto M^{0.271f_d - 0.06}$, indicating a turn-over of the concentration with significant decays (i.e. $f_d > 0.22$). We also investigated the variance of the logarithmic concentration of DDM haloes and found no obvious difference from the CDM case.

\subsection{The linear halo bias}\label{sec4-4}
\begin{figure}[tbp]
\centering
\includegraphics[width=1\textwidth]{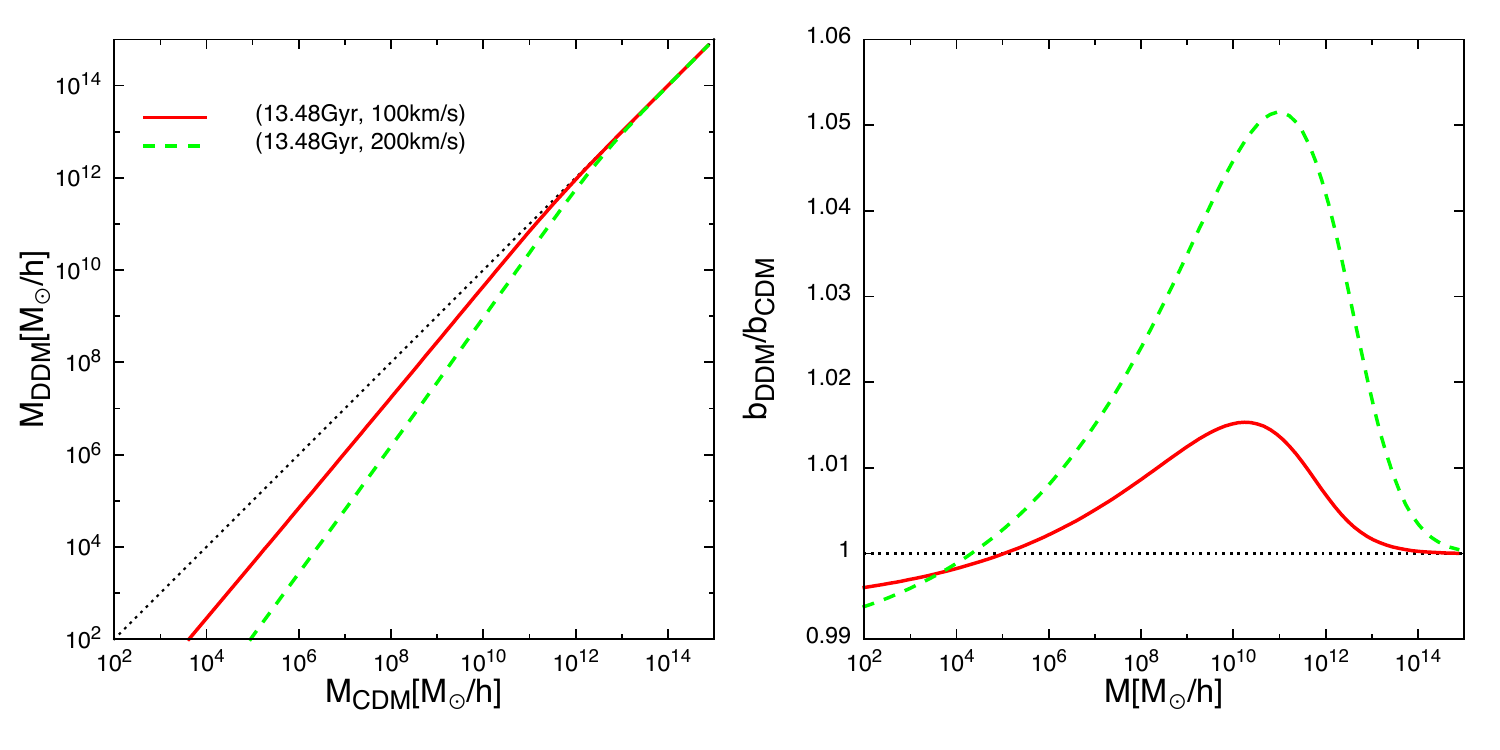}
\caption{\textit{Left}: The halo mass mapping for decay parameters $\tau=13.48$ Gyr and $V_k = 100, 200$ km/s at $z=0$. \textit{Right}: The ratio of the halo bias between DDM and CDM for the same decay parameters. The dotted lines in both panels represent the limit of infinitely small DDM suppression, where $M_{ \text{DDM}} = M_{ \text{CDM}}$.}
\label{fig-ddm-bias}
\end{figure}

The linear bias of CDM haloes is usually understood from the peak-background split \cite{pb-split}. For the ST formalism, the bias is \cite{st-mf}
\begin{equation}\label{eq32}
b_{ \text{CDM}} = 1 + \frac{q\nu^2 -1}{\delta_c(z)}+ \frac{2p}{\delta_c(z)\left[1+(q\nu^2)^p\right]},
\end{equation}
where $\nu$ depends on $M$ and the parameters $p$ and $q$ are the same as in Eq. (\ref{eq25}).  Because the DDM suppression is only important in late times, the statistics of the early density field is the same as CDM. However, due to the decays, a proto-region that ought to collapse to a CDM halo of mass $M_{ \text{CDM}}$ may now form a DDM halo of mass $M_{ \text{DDM}}$. The two haloes could occupy similar positions, which can also be seen in Fig. \ref{fig-cdm-ddm-wdm}. The DDM halo bias is then known after adapting a mass mapping between the CDM and DDM haloes. 

Physically, $M_{\text{DDM}}$ is expected to be smaller than $M_{\text{CDM}}$ through several ways of mass loss: (1) recoiling daughters can directly escape from shallow potentials; (2) the decays can reduce halo concentrations, which makes them more vulnerable to tidal stripping; (3) with mean halo density reduced, the virial radius would also shrink to enclose less mass. Making an optimistic assumption that the decay effects still preserve the number of haloes, we construct the mass mapping as
\begin{equation}\label{eq33}
\frac{\text{d}M_{ \text{DDM}}}{\text{d}M_{ \text{CDM}}} = \frac{n_{ \text{CDM}}(M_{ \text{CDM}}, z)}{n_{ \text{DDM}}(M_{ \text{DDM}}, z)},
\end{equation}
where $M_{ \text{DDM}}$ is a function of $M_{ \text{CDM}}$ and on the right hand side are the CDM and DDM mass functions. Because decays only modify the structures within non-linear scales for small $V_k$, the linear bias of DDM halo is therefore
\begin{equation}\label{eq34}
b_{ \text{DDM}}(M_{ \text{DDM}}) = b_{ \text{CDM}}(M_{ \text{CDM}}) .
\end{equation}

We solve the equations numerically by involving Eq. (\ref{eq28}).  In Fig. \ref{fig-ddm-bias}, we present the mass mapping and the bias for two sets of decay parameters. In the mass mapping, the DDM suppression is always stronger towards lower mass haloes. But this is not true for the bias as shown in the right panel, since the CDM bias is no longer sensitive to the mass difference in the low mass end. Instead, the largest difference appears in the medium mass range, where the DDM suppression is important and the CDM bias increases quickly with the halo mass. However, the overall difference is still small. As we will see in Fig. \ref{fig-ddm-roles}, it hardly contributes to the power suppression.

\subsection{Reconstruction of the DDM power suppression}\label{sec4-5}
\begin{figure}[tbp]
\centering
\includegraphics[width=0.7\textwidth]{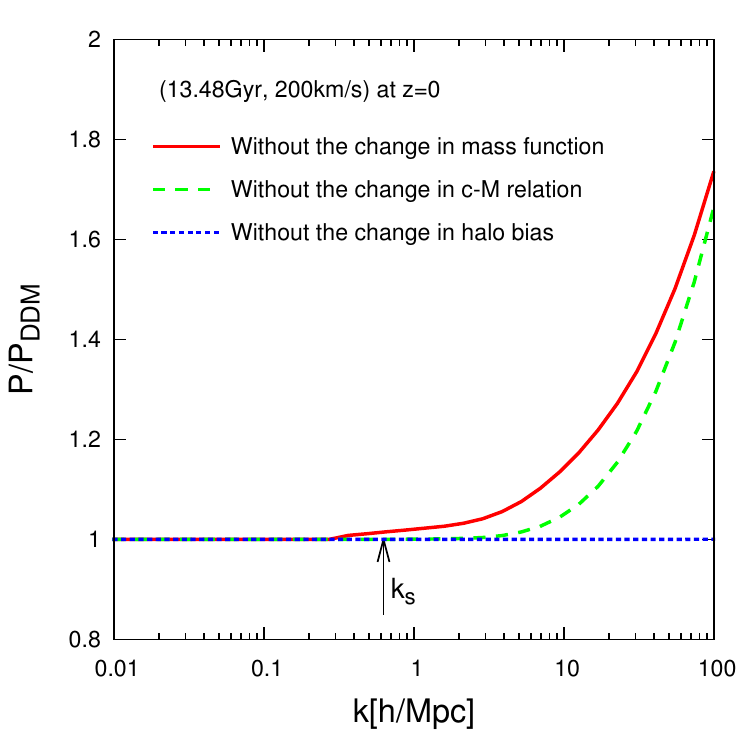}
\caption{Ratios of the DDM power without the changes in the mass function (red solid), c-M relation (green  dashed) and halo bias (blue dotted) to that including all the effects at $z=0$. The decay parameters here are $\tau = 13.48$ Gyr and $V_k = 200$ km/s. The arrow here points to the characteristic suppression scale $k_s$ as defined in Eq. (\ref{eq8}).}
\label{fig-ddm-roles}
\end{figure}

\begin{figure}[tbp]
\centering
\includegraphics[width=0.7\textwidth]{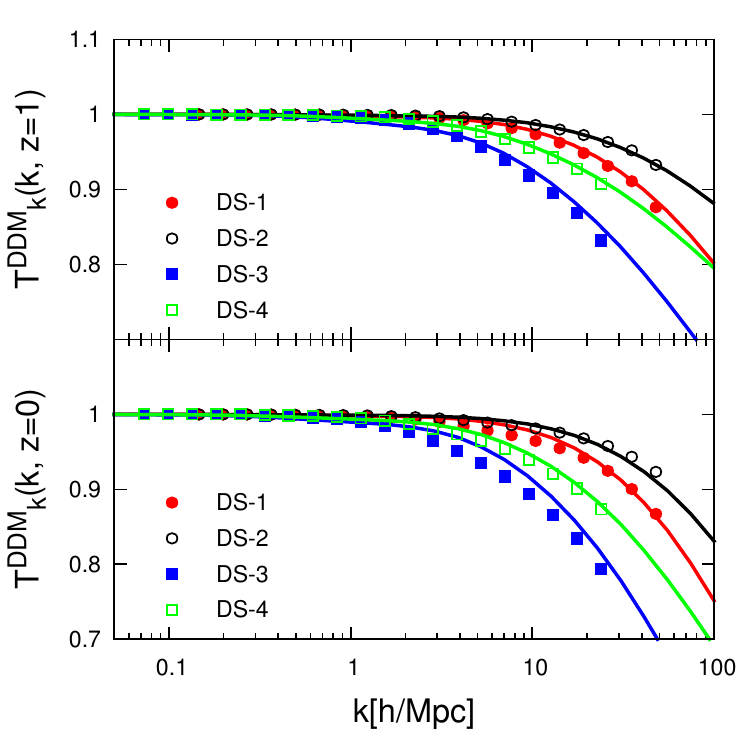}
\caption{Comparison of the modelled DDM transfer functions (lines) and the simulation data (points). The top and bottom panels refer to the results at $z=1$ and $0$. The DS-3 and DS-4 simulations of 100 Mpc/$h$ together with the DS-1 and DS-2 simulations of 50 Mpc/$h$ are used for the plot. }
\label{fig-ddm-transfer-function}
\end{figure}

Before applying the halo model we described, we also need to know the mass fraction of all haloes. For WDM, Ref. \cite{hm-2} suggested a physical cutoff mass $M_{ \text{cut}}$ according to the WDM free-streaming scale. This assumption can avoid the problems of ambiguous extrapolation of the mass function to the low mass end and also the numerical instabilities of Eq. (\ref{eq24}) when $M_{\text{cut}} \rightarrow 0$. However, we have found no sign of physical cutoff in the DDM mass function. However, as we show in Appendix \ref{app-b}, the small enough haloes could act exactly like the smooth component. We therefore argue that the cutoff mass is still appropriate for the DDM halo model calculation, but now it should refer to the mass scale below which the lower mass haloes and the real smooth component are indistinguishable. In this sense, the standard halo model and the halo model with smooth component are unified.

In Fig. \ref{fig-ddm-roles}, we firstly examine the roles of the DDM effects on the mass function, halo profile and halo bias on the final halo model power. Each line represents a result without certain modification of the halo model ingredient. We can see that the DDM reduction on the mass function is mostly important for the power suppression and also marks the characteristic scale. The change in the c-M relation starts to be important in smaller scales, while the effect of linear bias difference is negligible.

In Fig. \ref{fig-ddm-transfer-function}, we finally compare the calculated DDM transfer functions with the data from simulations. The halo model predictions describe well the data to the non-linear scales. The modelling seems even better if the DDM suppression is smaller. The largest mismatch is in simulation DS-3 at $z=0$ and the range $k \sim (2 - 10)$ $h$/Mpc with the relative error about 2\%. We also found that the suppression tails of different lifetimes do not overlap when the scale is normalized by $k_s$, meaning that the transfer function cannot be described by any function in the form $f(k/k_s, f_d)$. The reason is that the contributions from $V_k$ and $\tau$ are no longer separable for the suppression tails. We also want to draw the attention that the ingredients of the DDM halo model are summarized from the small suppression set simulations. A safe parameter  range to utilize these results is $V_k$ smaller than 200 km/s and $\tau$ lager than or at least the same order as the cosmic age. The large suppression set simulations are found to have big deviations from the semi-analytical predictions. But fortunately, they are also not consistent with observations.

\section{Discussions}\label{sec5}
Based on the previous results, we discuss (1) the constraints of the DDM parameter space,  (2) the more generalized DDM models with non-trivial mother particle initial fractions and (3) whether the DDM suppression can be a solution to the Planck Sunyaev-Zeldovich and primary CMB disagreement.

\subsection{The Lyman-$\alpha$ constraints}\label{sec5-1}
\begin{figure}[tbp]
\centering
\includegraphics[width=0.7\textwidth]{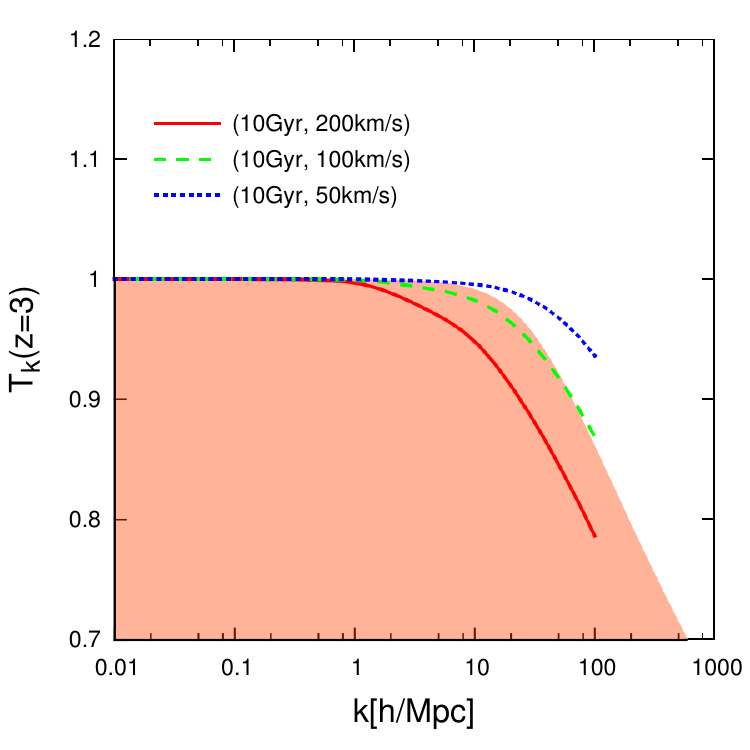}
\caption{The transfer functions of three DDM models compared with a 2 keV WDM at $z=3$. The lines correspond to the DDM models of $\tau = 10$ Gyr and $V_k = (50, 100, 200)$ km/s, which will cause less flux power of Lyman-$\alpha$ if the suppression is within the colored region corresponding to that of WDM.}
\label{fig-ddm-wdm-demo}
\end{figure}

\begin{figure}[tbp]
\centering
\includegraphics[width=0.7\textwidth]{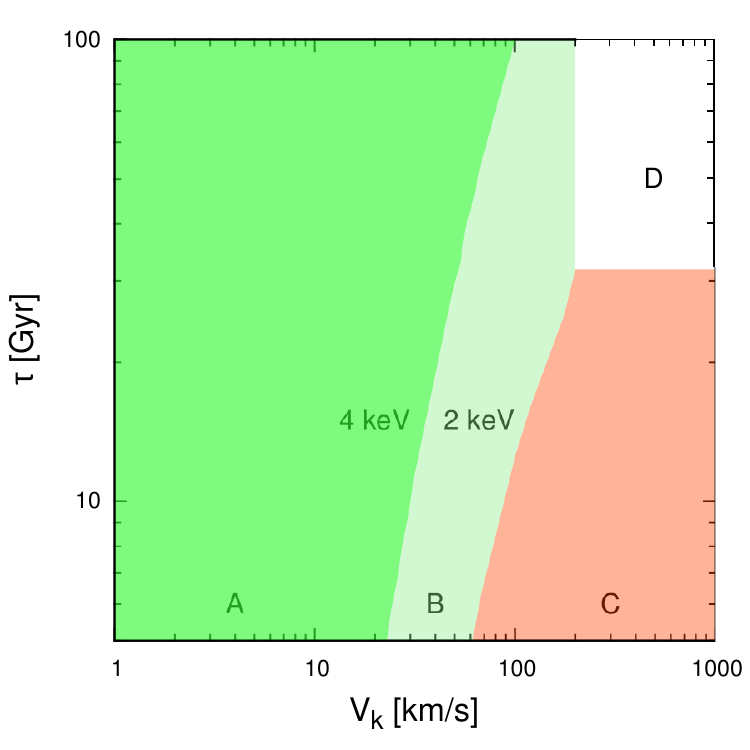}
\caption{The constraints on the DDM parameter space by translating the Lyman-$\alpha$ constraints on thermal WDM.  The DDM halo model is applied in the range $\tau \ge 5$ Gyr and $V_k \le 200$ km/s. The allowed parameter space include region A for the 4 keV WDM case, and will extend to include region B for the 2 keV case. Region C is ruled out for our consideration and region D is not explored as $k_s$ there may touch linear scales.}
\label{fig-ddm-constraints-special}
\end{figure}

We consider the Lyman-$\alpha$ constraints on the DDM parameters. The Lyman-$\alpha$ forest is the neutral hydrogen absorption lines in the spectra of distant quasars (QSOs). As the hydrogen clouds are tracing the density perturbations, the flux spectrum $P_{F}(k,z)$ is imprinted with the information of the underlying density field at medium redshift $z \sim (2 - 4)$ and on scales $k \sim (0.1- 10) h/\text{Mpc}$ that have not been fully contaminated by non-linear evolution. Usually, the flux bias function $b_{F}(k, z) = P_{F} (k,z)/P(k, z)$ that relates the flux power to the real density power has to be understood before interpreting the data. However, it has a complicated dependence on the cosmological parameters, the initial power spectrum as well as the parameters of the baryon physics \cite{Lyman-2, Lyman-3, Lyman-4}. A large number of hydrodynamical simulations are in principle needed to do so.  Here we simply assume that the flux bias function is unchanged for the WDM and DDM models at the redshift and scales relevant to the Lyman-$\alpha$ observations. We can then translate the Lyman-$\alpha$ limits on WDM to DDM.  Considering the their opposite evolution tendencies, this assumption might underestimate the true flux power in the DDM models and make the DDM constraints conservative. 

The preferred DDM parameters are those that cause less suppression on the CDM power than that from the lower mass limit of WDM particles. The Lyman-$\alpha$ forest has been shown to be sensitive to the WDM mass with the lower mass limit between $(2-4)$ keV for thermal WDM particles \cite{Lyman-1, Lyman-5, Lyman-6, Lyman-7}, where the data set probing higher redshift usually concludes in higher mass. In Fig. \ref{fig-ddm-wdm-demo}, we give an example of the translation by showing the power suppression for a 2 keV WDM and several DDM models at $z=3$. The transfer functions of the DDM models are calculated from the DDM halo model. The suppression of the WDM is denoted as the color region, where we have adopted a well calibrated fitting formula for the WDM transfer function from Ref. \cite{Lyman-8}.  The DDM models within the color region have larger suppression than the WDM model and are thus not favoured. We then survey the parameter space with $\tau \ge 5$ Gyr and $V_k \le 1000$ km/s. The results are shown in Fig. \ref{fig-ddm-constraints-special}, where the parameter space is divided into four regions (A, B, C, and D). Region A is conservative and allowed by the 4 keV WDM limit. Region B is in between the 4 keV and 2 keV limits. Region C is ruled out, but region D is still uncertain because the accuracy of the DDM halo model is inadequate. At $\tau = 13.48$ Gyr, the constraints of the recoil velocity are that $V_k$ should be smaller than $(35-105)$ km/s. While at $\tau = 10$ Gyr, the constraints are $V_k$ less than $(31-88)$ km/s. This result is consistent with a recent direct fitting of DDM models to the Lyman-$\alpha$ data in Ref. \cite{ddm-wang-2}, where they have concluded that for $\tau \le 10$ Gyr, $V_k$ is smaller than $(30-70)$ km/s.

%Using the same data set, the constraint on WDM mass is about 2.12 keV from Boyarsky et al. (2008).

\subsection{More generalized initial condition}\label{sec5-2}

\begin{figure}[tbp]
\centering
\includegraphics[width=0.7\textwidth]{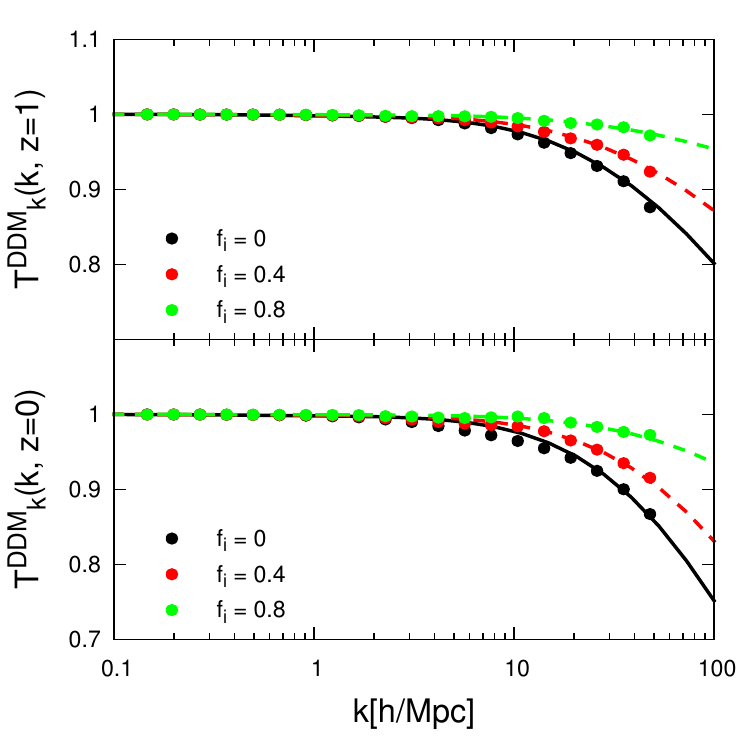}
\caption{Comparison of the halo model predictions by replacing $f_d$ with $f_g$ and the transfer functions measured from the DDM simulations with $f_i=$0, 0.4 and 0.8. The decay parameters here are $\tau = 13.48$ Gyr and $V_k = 100$ km/s.}
\label{fig-ddm-initial-fraction}
\end{figure}

\begin{figure}[tbp]
\centering
\includegraphics[width=0.7\textwidth]{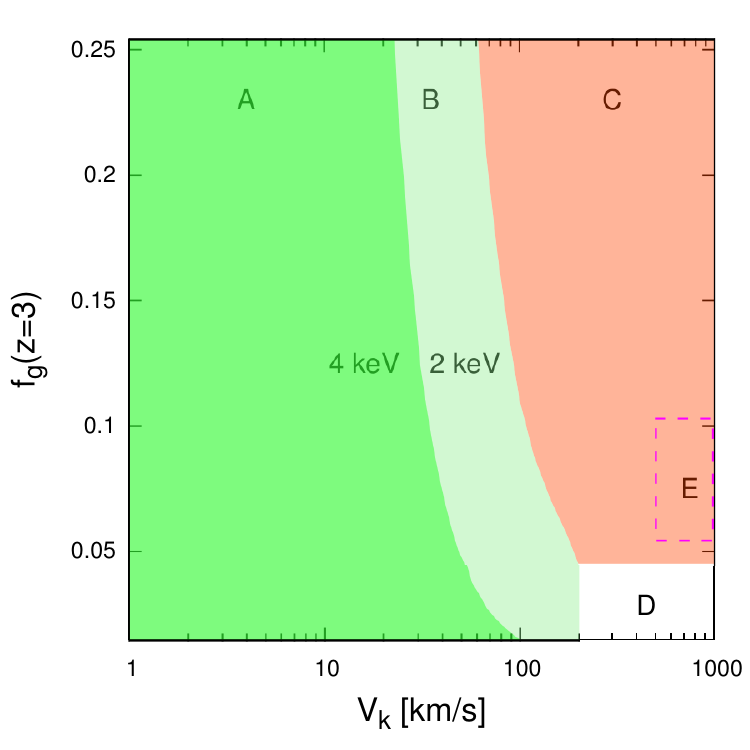}
\caption{The generalized constraints of DDM models on global decayed fraction and recoil velocity. The regions A to D represent the same regions as those in Fig. \ref{fig-ddm-constraints-special}. The region E bounded with the magenta dashed lines corresponds to the parameter space of resolving the Planck disagreement on the cluster number with $f_i = 0$.}
\label{fig-ddm-constraints-general}
\end{figure}

In all previous studies of DDM, the mother particles $ddm$ are assumed to make up all the matter component initially. We relax this limit by considering that the $ddm$ only contributes to parts of the initial mass,  leaving the other mass as stable matter with the fraction
\begin{equation}
f_i = 1- \frac{\Omega_{\text{ddm}}(z_i)}{\Omega_{m}(z_i)}
\label{initial-daughter-fraction}
\end{equation}
at $z_i$. The stable matter might be the daughter particles $dm$ or baryons or even other type of CDM. The first possibility may be realized if $ddm$ and $dm$ are both WIMPs and are thermally produced with comparable annihilation cross sections. We expect the stable and unstable components are uniformly mixed initially. Effectively, if a global decayed fraction $f_g = (1-f_i)f_d$ can be mimicked by an other lifetime $\tau_{\text{eff}}$ with $f_i=0$ such that
\begin{equation}\label{eq36}
f_d(\tau_{\text{eff}}, z) = f_g(\tau, z),
\end{equation}
the structure formation of the two cases should be exactly the same. However, the solution of Eq. (\ref{eq36}) is time dependent in general, unless in the very long lifetime limit  ($\tau \gg H_{0}^{-1}$), the effective lifetime approaches a constant
\begin{equation}\label{eq38}
\tau_{\text{eff}} \rightarrow \frac{\tau}{1-f_i}, 
\end{equation}
implying that $f_i$ and $\tau$ can be degenerate. 

For the normal situation with $\tau \sim H_{0}^{-1}$, non-zero $f_i$ will cause a different accumulation of the decay produced daughters, which can be studied by changing the $\eta$ of the N-body algorithm to
\begin{equation}\label{eq39}
\eta(T_1) =  \left[1-\exp \left(-\frac{\ln 2}{\tau}\cdot \frac{T_s}{f_s} \right) \right] \cdot \frac{(1-f_i)}{(1-f_i) + f_i \exp \left(\frac{\ln 2}{\tau}\cdot T_1\right)}.
\end{equation}
We redo the simulation DS-1 of 50 $h$/Mpc with the new ratio and consider $f_i=40\%$ and $80\%$ that correspond to $f_g = 30\%$ and $10\%$ at $z=0$. Since it is the new born daughters that cause the suppression, we also generalize the DDM halo model by replacing $f_d$ with $f_g$. The transfer functions from the new simulations and new halo model are compared in Fig. \ref{fig-ddm-initial-fraction}. The good agreement shows that the global decayed fraction $f_g$ is the right parameter to describe the generalized DDM models. Also notice that we previously only summarized the DDM halo model ingredients from the simulations with the decayed fraction of 30\% and 50\% at $z=0$; the match to the new decayed fraction of 10\% also demonstrates the accuracy of halo model to the small suppression end. Another point to notice is that to have the same $f_g$ at a fixed redshift, there is a degeneracy in $f_i$ and $\tau$. However, the degeneracy should be broken if the suppression is examined at other redshifts.

Based on Fig. \ref{fig-ddm-constraints-special}, we constrain the more generalized DDM models in Fig. \ref{fig-ddm-constraints-general}. Fig. \ref{fig-ddm-constraints-general} also allows us to make a first order correction to the DDM constraints with baryonic matter, which has been neglected previously. By assuming that baryons honestly trace the undecayed mass, their presence is equivalent to an effective $f_i = \Omega_b / \Omega_m =  16.3 \%$.  The DDM constraints with all initial dark matter as mother particles at $\tau = 13.48$ Gyr are broadened to $V_k$ less than $(37-120)$ km/s.

\subsection{Resolving the Planck disagreement?}\label{5-3}

\begin{figure}[tbp]
\centering
\includegraphics[width=0.7\textwidth]{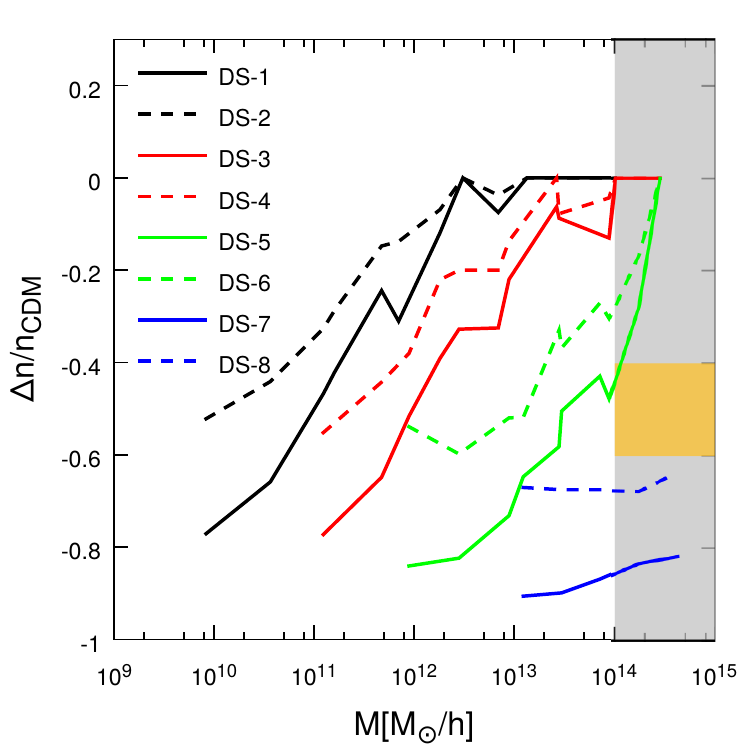}
\caption{Halo number density differences of DDM normalized by the number density of CDM measured in simulations with $\Delta=500$ at $z=0$. The grey region shows typical cluster mass range. The yellow region represents the suppression of DDM needed to resolve the Planck disagreement. DDM simulations that have different box sizes in Table \ref{table1} are combined to make the plot.}
\label{fig-ddm-cluster}
\end{figure}

The Planck 2013 results have reported the constraints of the cosmological parameters using the number counts of the Sunyaev-Zeldovich effect selected clusters \cite{planck-2013}. The interpolated parameters $\sigma_8$ and $\Omega_m$ are found to differ from those derived from the primary CMB temperature anisotropies, which is also indicated by the latest data release \cite{planck-2015}. With a reasonable bias of the measured cluster mass to the real mass, it could lead to the number of predicted clusters from CMB two times larger than the observed. 

DDM may provide a natural explanation for this disagreement, as it can reduce the cluster number in the late evolution without interfering the CMB.  Such possibility was firstly proposed in Ref. \cite{aoyama}, but only linear perturbations have been done. With our N-body simulations, we revisit the DDM suppression on the mass function with a higher $\Delta =500$ for the cluster convention \cite{planck-2013}. Focusing on the cluster mass range, we see in Fig. \ref{fig-ddm-cluster} that the small suppression set (DS-1 to DS-4) makes no difference to CDM. The preferred DDM suppression then should reduce the cluster number by more than those of DS-5 and DS-6 but less than those of DS-7 and DS-8. For $\tau \sim H_{0}^{-1}$, this parameter space is shown as region E in Fig. \ref{fig-ddm-constraints-general}, which however is deeply inside the Lyman-$\alpha$ ruled-out region C. The reason is that to have such a large decrease of high mass haloes, smaller haloes are already suppressed much more. The $f_i$ here is zero. Higher $f_i$ would not change the conclusion, since that will require even faster decays and move the region E upwards. Also notice that our simulations do not have the same cosmology as Planck measured. However, the Planck cosmology will produce even more high mass clusters.  We expect that the region E is also conservative for the exact Planck cosmological parameters. 

The failure of the DDM models suggests that the late-time suppression shall only occur in high mass objects. In this sense, mechanisms like the AGN feedback \cite{agn-feed-1, agn-feed-2} might be more plausible to resolve the disagreement.

\section{Summary and conclusion}\label{sec6}
In this paper, we studied the cosmological structure evolution in the non-relativistic and long-lifetime DDM models. The decay mechanism brings unique features to the structure evolution, with the recoil velocity $V_k$ determining a characteristic suppression scale and the lifetime $\tau$ regulating the time when the suppression is important. Intrinsically different from the WDM, the structures in DDM models are more CDM like in early times. We argue that DDM models could more easily cause suppression in the late universe while being consistent with the high redshift observations, such as the reionization and Lyman-$\alpha$ forest.  

In particular, we considered the DDM models with two-body decay and only one type of massive daughter. The two possible cases (Model A and B) were shown to be identical in the structure formation. Using N-body simulations, we solved the coupled equations that govern the DDM structure evolution from the first principle. These cosmological simulations are needed to understand the effects of DDM, especially on non-linear scales. In the analytical aspect, we proposed empirical functions parameterized by the characteristic mass $M_f$ and decayed fraction $f_d$, which are again functions of the decay parameters and redshift, to describe the DDM suppression on the mass function and the halo concentration. This also leads to accurate reconstruction of the non-linear power transfer function of DDM to CDM in the framework of halo model. The consequence of these efforts is that using the analytical predictions the decay parameter space can be explored far more than the few simulated points.

Additionally, we translated the constraints of WDM mass from Lyman-$\alpha$ forest to those of the DDM parameters. The results were shown in Fig. \ref{fig-ddm-constraints-special}. For decay models with $\tau \sim H_{0}^{-1}$, we found the $V_k$ should be smaller than 105 km/s or 35 km/s for more conservative consideration. The DDM models are also generalized to arbitrary initial fractions of the mother particles. We found that the halo model is still valid after replacing $f_d$ with the global decayed fraction $f_g = (1-f_i)f_d$. Constraints were also made for the generalized DDM models as shown in Fig. \ref{fig-ddm-constraints-general}. Using the constraints, we also demonstrated that the DDM models are unlikely to resolve the disagreement on the cluster numbers from the Planck SZ survey and primary CMB prediction without violating the Lyman-$\alpha$ limits.

Through the study, we have shown that DDM is rich in phenomenology and its ability of reducing high density and suppressing the formation of small structures has been revealed. In future work, we will substitute DDM for CDM to other dark matter related observations, such as weak lensing, HI surveys of the galaxy velocity function and dark matter detections. We also expect higher resolution simulations of our algorithm to explore more details of DDM structures on smaller scales \footnote{It would be interesting to compare the profiles and mass function of subhaloes with the zoom-in simulations of Wang et al. \cite{ddm-wang-3}, where they have applied the Peter's algorithm.}.  Together, we may have an answer of whether DDM can be a better alternative of CDM.
 
\appendix

\section{Convergence tests of the simulations}
\subsection{Test the artificial decay parameters}\label{app-a-1}
\begin{figure}[tbp]
\centering
\includegraphics[width=1\textwidth]{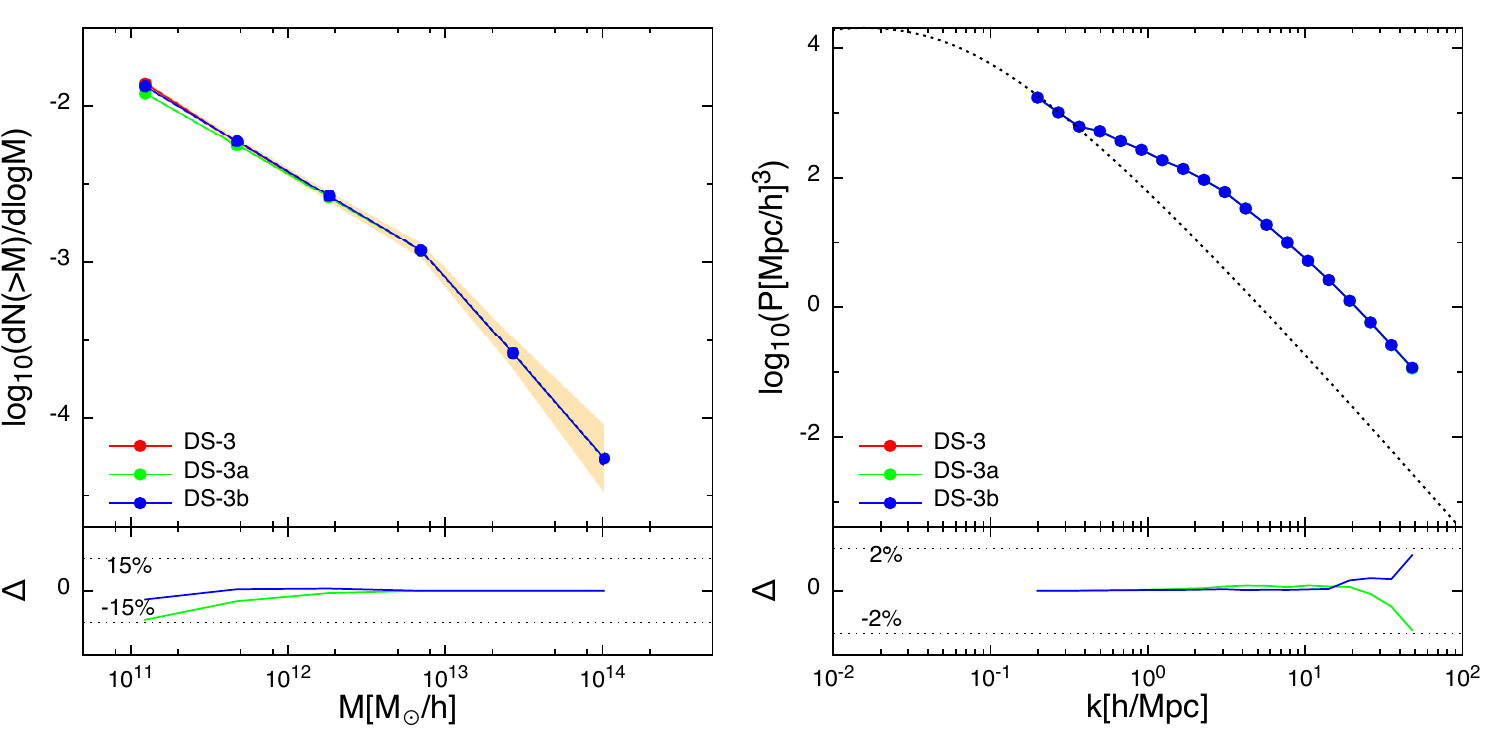}
\caption{Consistency tests of the artificial decay parameters, where DS-3a and DS-3b have either high value in $f_s$ and $N_s$ than DS-3 of the same simulation box. \textit{Left side}: Their mass functions at $z=0$. The colored region represents the errors of DS-3. \textit{Right side}: Their power spectra at $z=0$ with the dashed line showing the linear power.  In both sides, the green and blue lines in the bottom panels represent the differences of DS-3a and DS-3b with respect to DS-3.}
\label{fig-consistent-tests}
\end{figure}

 Physical properties extracted from the DDM simulation should depend on the decay parameters ($\tau$ and $V_k$) rather than the artificial simulation parameters ($f_s$ and $N_s$).  As listed in Table \ref{table1}, two test simulations DS-3a and DS-3b are performed, which have either higher $f_s$ or $N_s$ than DS-3.  We examine the convergence of the simulations by comparing their mass functions and power spectra at $z=0$. As shown in the left panel of Fig. \ref{fig-consistent-tests}, DS-3a and DS-3b have the same number density of haloes as DS-3 in the high mass end. Meanwhile, DS-3b is closer to DS-3 than DS-3a in the low mass end, indicating that the parameter $N_s$ is more easily converged than $f_s$. However, the overall differences are still less than 15\%. The right panel shows an better consistency in the power spectra, where the largest difference is less than 2\% even on highly non-linear scales. Since changing $N_s$ and $f_s$ will let the simulations run with completely different particles, we believe that the choice of $f_s=10$ and $N_s = 1$ already makes the simulations converged, and we use them as the default parameters for other DDM simulations.

\subsection{Test the DDM suppression with box sizes} \label{app-a-2}
\begin{figure}[tbp]
\centering
\includegraphics[width=0.7\textwidth]{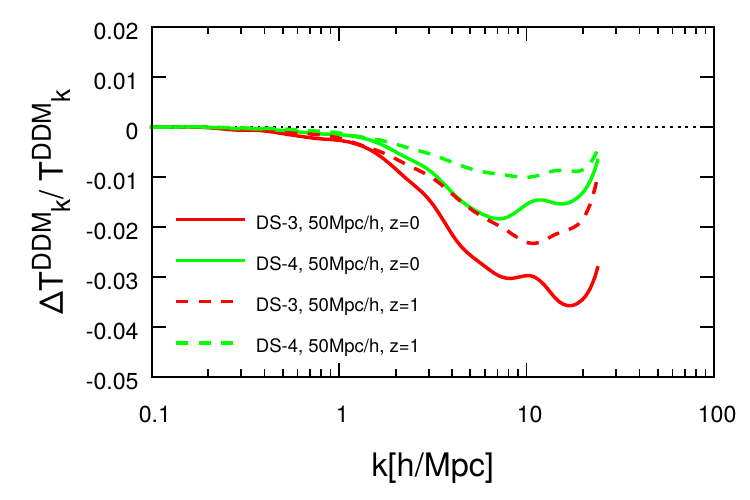}
\caption{Consistency tests of the DDM transfer functions with the simulation boxes. The relative differences of the transfer functions from boxes of 50 Mpc/$h$ are compared with the results from boxes of 100 Mpc/$h$.  DDM parameters of DS-3 and DS-4 are shown with the red and green lines, and redshifts at $z=0$ and 1 are further represented with the solid and dashed styles.}
\label{fig-boxes-tests}
\end{figure}
In Fig. \ref{fig-boxes-tests}, we test the convergence of the DDM power transfer functions defined in Eq. (\ref{eq9}) of DS-3 and DS-4 with the simulation box sizes. The results of 50 Mpc/$h$ are compared to the results of 100 Mpc/$h$. The largest differences occur in DS-3 at $z=0$, which is about 3.5\% at $k \sim 17$ $h$/Mpc before getting smaller at higher $k$. The differences are also redshift and DDM suppression dependent, with better consistency at high redshift and with smaller DDM suppression.

\section{Equivalence of small haloes and smooth component in the halo model}\label{app-b}
The standard halo model can be expressed as
\begin{equation}\label{eq40}
P(k) = P_{ \text{2H}}(k)+P_{ \text{1H}}(k),
\end{equation}
where the two- and one-halo terms are
\begin{equation}\label{eq41}
P_{ \text{2H}}(k) = \frac{1}{\bar{\rho}^2_m}P_{\text{lin}}(k)\left[ \int ^{\infty}_{0} dM M b_1(M) n(M) \tilde{u}(k|M) \right]^2
\end{equation}
and
\begin{equation}\label{eq42}
P_{ \text{1H}}(k) = \frac{1}{\bar{\rho}^2_m}\int^{\infty}_0 dM n(M) M^2 \tilde{u}^2(k|M).
\end{equation}
The explicit form of the mass filter of a NFW halo is
\begin{equation}\label{eq43}
\begin{split}
\tilde{u}(k|M) = & \frac{4\pi\rho_s r_s^3}{M} \lbrace \sin(kr_s)\left[ \text{Si}[(1+c)kr_s] - \text{Si}(kr_s)\right] - \frac{\sin(ckr_s)}{(1+c)kr_s} \\
                          &+ \cos(kr_s)\left[\text{Ci}[(1+c)kr_s] - \text{Ci}(kr_s) \right] \rbrace,
\end{split}
\end{equation}
where $\text{Si}(x) = \int^{x}_0 \sin(t)/t \text{d}t$ and $\text{Ci}(x) = -\int^{\infty}_{x} \cos(t)/t \text{d}t$. Eq. (\ref{eq43}) can reach the limit
\begin{equation}\label{eq44}
\lim_{kr_s \rightarrow 0}\tilde{u}(k|M) =\frac{4\pi\rho_s r_s^3}{M} \left[\ln(1+c) - \frac{c}{1+c}\right] = 1,
\end{equation}
when $kr_s \ll 1$, where the last equality is the mass conservation equation of the NFW profile. The limit can be approached on large enough scales or in haloes of small enough $r_s$. Demanding that the two-halo term equals the linear power on large scales, Eq. (\ref{eq41}) and Eq. (\ref{eq44}) lead to a non-trivial bias mass relation \cite{seljak}
\begin{equation}\label{eq45}
\int^{\infty}_{0}dM M b_1(M)n(M) = \bar{\rho}_m.
\end{equation}

\begin{figure}[tbp]
\centering
\includegraphics[width=0.7\textwidth]{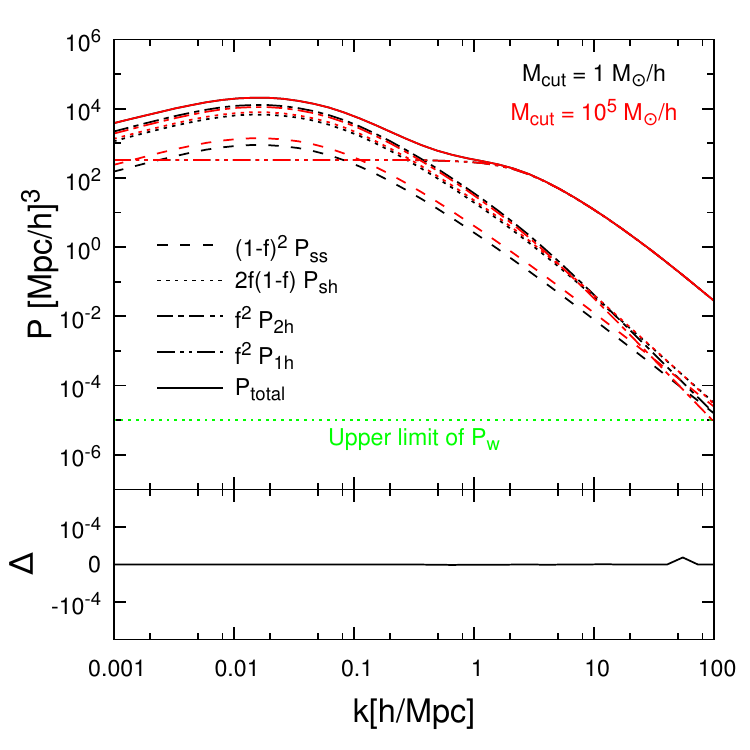}
\caption{The standard halo model power calculated using different cutoff mass. In the top panel, the dashed, dotted, dashed-dotted and the dashed-double-dotted lines correspond to the terms in Eq. (\ref{two-halo-term-smooth}) and (\ref{one-halo-term-smooth}). The solid lines are the sum of the terms.  The results from $M_{ \text{cut}}=10^5 \text{M}_{\odot}/h$ and $1 \text{M}_{\odot}/h$ are represented with the red and black colors, respectively. The green dotted line shows the upper limit of the white noise of $M_{ \text{cut}}=10^5 \text{M}_{\odot}/h$. In the bottom panel, we plot the fractional difference of the total power of $M_{ \text{cut}}=10^5 \text{M}_{\odot}/h$ with respect to that of $M_{ \text{cut}}=1 \text{M}_{\odot}/h$.}
\label{fig-ddm-cutoff}
\end{figure}

Assuming a cutoff  mass $M_{\text{cut}}$ below which the largest wave number of interest satisfies $k_{ \text{max}}r_s(M_{ \text{cut}}) \ll 1$ \footnote{The DDM models can also satisfy the inequality, because $r_s \propto r_{vir} /c \propto M^{0.72-0.271f_d}$ by extending Eq. (\ref{eq31}) to the low mass end.}, we have
\begin{equation}\label{eq46}
\begin{split}
\int^{\infty}_{0}dM M b_1(M)n(M)\tilde{u}(k|M) =  &\int^{M_{ \text{cut}}}_0 dM M n(M) b_1(M)\\
 &+ \int^{\infty}_{M_{ \text{cut}}} dM M n(M) b_1(M)\tilde{u}(k|M).
\end{split}
\end{equation}
After defining the mass fraction for haloes larger than $M_{ \text{cut}}$ as
\begin{equation}\label{eq47}
\begin{split}
f & = \frac{1}{\bar{\rho}_m}\int^{\infty}_{M_{ \text{cut}}}dM M n(M) \\
 & = \frac{\bar{\rho}_h}{\bar{\rho}_m}, 
\end{split}
\end{equation}
the first term on the right hand side of Eq. (\ref{eq46}) can be rewritten as
\begin{equation}\label{eq48}
\int^{M_{ \text{cut}}}_0 dM M n(M) b_1(M) = b_s (1-f)\bar{\rho}_m.
\end{equation}
Combining Eq. (\ref{eq48}) and Eq. (\ref{eq45}), we have
\begin{equation}\label{eq49}
\begin{split}
b_s  & = \frac{1}{(1-f)\bar{\rho}_m}\left[\bar{\rho}_m - \int^{\infty}_{M_{ \text{cut}}} dM M b_1(M) n(M) \right]\\
        & = \frac{1-fb^{ \text{eff}}}{1-f},
\end{split}
\end{equation}
with the form of the effective bias as
\begin{equation}\label{eq50}
b^{ \text{eff}} = \frac{1}{\bar{\rho}_h}\int^{\infty}_{M_{ \text{cut}}}dM M n(M) b_1(M).
\end{equation}
Inserting Eq. (\ref{eq46}) and Eq. (\ref{eq48}) back to Eq. (\ref{eq41}), we have the two-halo term
\begin{equation}
\begin{split}
P_{ \text{2H}} = & (1-f)^2b_s^2P_{\text{lin}} + 2(1-f) f\frac{b_sP_{\text{lin}}}{\bar{\rho}_h}\int^{\infty}_{M_{ \text{cut}}}dM M n(M)b_1(M)\tilde{u}(k|M)  \\
                                              &+f^2\frac{P_{\text{lin}}}{\bar{\rho}_h^2}\left[\int^{\infty}_{M_{ \text{cut}}}dM M n(M)b_1(M)\tilde{u}(k|M)\right]^2 \\
                                               = &(1-f)^2P_{ss} +2(1-f)fP_{sh}+f^2P_{2h}.                                               
\end{split}
\label{two-halo-term-smooth}    
\end{equation}
Using the same strategy, the one-halo term is separable as
\begin{equation}
\begin{split}
P_{ \text{1H}} &= f^2 \frac{1}{\bar{\rho}_h^2}\int^{\infty}_{M_{ \text{cut}}}dM n(M)M^2\tilde{u}^2(k|M) +\frac{1}{\bar{\rho}^2_m}\int^{M_{ \text{cut}}}_0 dM n(M) M^2 \\
                                          &= f^2P_{1h}+P_{w}.                       
\end{split}      
\label{one-halo-term-smooth}                              
\end{equation}
Therefore, we have shown that the standard halo model and the halo model with unbounded mass are equivalent in calculation, except for a negligible white-noise term
\begin{equation}
\begin{split}
P_{w} & = \frac{1}{\bar{\rho}^2_m}\int^{M_{ \text{cut}}}_0 dM n(M) M^2 \\
               & < \left[ \frac{1}{\bar{\rho}_m}\int^{M_{ \text{cut}}}_0 dM n(M) M \right]  \frac{M_{ \text{cut}}}{\bar{\rho}_m} \\
               & = (1-f) \frac{M_{ \text{cut}}}{\bar{\rho}_m} \\
               & \leq \frac{M_{ \text{cut}}}{\bar{\rho}_m}.
\end{split}
\end{equation}
The white noise can be smaller than $10^{-5}$ (Mpc/$h$)$^{3}$ if $M_{\text{cut}} < 10^5 \text{M}_{\odot}/h$. We present a test in Fig. \ref{fig-ddm-cutoff} by calculating the standard halo model with different cutoff mass. The good consistency shows that haloes smaller $10^5 \text{M}_{\odot}/h$ are already indistinguishable from the smooth component for $k_{\text{max}} < 100$ $h$/Mpc.

\acknowledgments
We thank Yipeng Jing and K. Dolag for discussions. Dalong Cheng would like to thank the CAA of SJTU for hospitality, where parts of the work have been done. We thank the anonymous referee for the improvements of the numerical details of the paper and also the ITSC of the Chinese University of Hong Kong for providing its cluster for computations. This work is partially supported by grants from the Research Grant Council of the Hong Kong Special Administrative Region, China (Project Nos. 400805 and 400910) and a Direct Grant from the Chinese University of Hong Kong.

\end{document}